\pdfoutput=1
\documentclass{JINST}

\usepackage{subfig}
\usepackage{amsmath,mathtools}
\usepackage[stretch=10,final]{microtype}

\DeclareCaptionLabelSeparator*{rperiod}{. }
\title{Target Mass Monitoring and Instrumentation in the Daya Bay Antineutrino Detectors}

\author{
	H.\,R.~Band$^a$,
	J.\,J.~Cherwinka$^b$,
	L.\,S.~Greenler$^b$,
 	K.\,M.~Heeger$^a$,
	P.~Hinrichs$^{a}$\thanks{Corresponding author.},
	L.~Kang$^{c}$,
	C.\,A.~Lewis$^{a}$,
	S.\,F.~Li$^{c}$,
	S.\,X.~Lin$^{c}$,
	M.\,C.~McFarlane$^{a}$,
	W.~Wang$^{da}$,
	D.\,M.~Webber$^a$\thanks{Corresponding author.},
    Y.\,D.~Wei$^{c}$,
	T.\,S.~Wise$^a$,
	Q.~Xiao$^b$,
	L.~Yang$^{c}$,
	Z.\,J.~Zhang$^{c}$ \\
\llap{$^a$}Department of Physics, University of Wisconsin--Madison,\\
  Madison, WI 53706, U.S.A.\\
\llap{$^b$}Physical Sciences Laboratory, University of Wisconsin--Madison,\\
  Stoughton, WI 53589, U.S.A.\\
\llap{$^c$}School of Electronic Engineering, Dongguan University of Technology,\\
  Dongguan 523\,808,  China\\
\llap{$^d$}Department of Physics, College of William \& Mary,\\
Williamsburg, VA 23187, U.S.A.\\
\llap{$^*$}E-mail: \email{phinrichs@wisc.edu}\\
\llap{$^\dagger$}E-mail: \email{dwebber@hep.wisc.edu}}

\abstract{%
The Daya Bay experiment measures $\sin^2 2\theta_{13}$ using functionally identical antineutrino detectors located at distances of 300 to 2000 meters from the Daya Bay nuclear power complex.
Each detector consists of three nested fluid volumes surrounded by photomultiplier tubes.
These volumes are coupled to overflow tanks on top of the detector to allow for thermal expansion of the liquid.
Antineutrinos are detected through the inverse beta decay reaction on the proton-rich scintillator target.
A precise and continuous measurement of the detector's central target mass is achieved by monitoring the the fluid level in the overflow tanks with cameras and ultrasonic and capacitive sensors.
In addition, the monitoring system records detector temperature and levelness at multiple positions.
This monitoring information allows the precise determination of the detectors' effective number of target protons during data taking.
We present the design, calibration, installation and in-situ tests of the Daya Bay real-time antineutrino detector monitoring sensors and readout electronics.}


\def\deg{\ensuremath{^\circ}}
\def\degC{\ensuremath{\deg\mathrm{C}}}
\def\ohm{\ensuremath{\Omega}}
\def\unit#1{\ensuremath{\;\mathclose{\mathrm{#1}}}}
\newcommand{\nuclide}[3][]{\ensuremath{\prescript{#3}{#1}{\mathrm{#2}}}}

\def\url#1{\href{#1}{\texttt{#1}}}

\def\citefillingpaper{~\cite{fillingpaper}}
\let\citefillingpaper\relax

\begin{document}



\section{Introduction}

\begin{figure}[b]\hfil
\includegraphics[height=57mm]{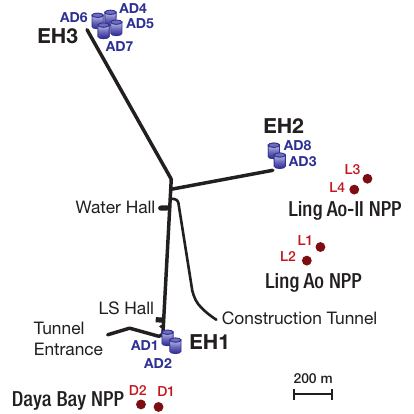}\hfil%
\caption{Site layout of the Daya Bay Reactor Neutrino Experiment. Six reactor cores are shown, along with eight locations for antineutrino detectors.}\label{fig:DYBsite}
\end{figure}

The goal of the Daya Bay reactor neutrino experiment is to make a measurement of the neutrino mixing angle $\sin^2 2\theta_{13}$ to a precision of 0.01 at 90\% confidence within three years of running \cite{Guo:2007ug}. The experiment has already released its first measurements of $\sin^2 2\theta_{13}$ \cite{DYBprl,DYBcpc}, with improved results to follow. Daya Bay is one of a new generation of reactor antineutrino disappearance experiments with near and far detector pairs located at kilometer-scale baselines from large nuclear power complexes. The experiment consists of eight antineutrino detectors (ADs) installed or under construction at three locations near the six-reactor Guangdong Nuclear Power Plant complex, located near Hong Kong and Shenzhen, China. The site layout is shown in Figure~\ref{fig:DYBsite}.  An accurate measurement of $\theta_{13}$ is of great importance to the particle physics community. Its value constrains many models of electroweak-sector physics and, prior to the first results from Daya Bay, had not been measured with high precision. Daya Bay's first experimental results \cite{DYBprl, DYBcpc, DYBad12} represent major progress toward the experiment's long-term goals and are a significant milestone in neutrino physics.

Like many other neutrino experiments, the Daya Bay antineutrino detectors (ADs) rely on the inverse beta-decay (IBD) reaction to detect antineutrinos:
\[\bar{\nu}_e + p \rightarrow e^+ + n.\]
The basic IBD reaction has an energy threshold of $1.806\unit{MeV}$ \cite{vogelIBD}; detectors using the IBD reaction are not sensitive to antineutrinos with energies below this threshold. The IBD cross-section is about $10^{-42}\unit{cm^2}$ at $E_{\bar\nu}=5\unit{MeV}$ \cite{vogelIBD}, necessitating a very sensitive detector. IBD on free protons is an especially desirable reaction for detector use as its signature is both clean and distinctive. The positron released during the event annihilates quickly on a detector electron, producing annihilation radiation; this is the ``prompt signal'' of an IBD event. The released neutron can also be detected if additional target nuclei that have both a high neutron capture cross-section and emit a clean energy signal on neutron capture are present in the detector. Natural gadolinium, containing \nuclide{Gd}{157} and \nuclide{Gd}{155}, is used for this purpose in the Daya Bay detectors. Both nuclei have very large neutron capture cross-sections and clean neutron capture signatures. The photons released following a neutron capturing on Gd in the detector form the ``delayed signal'' of an IBD event. This combination of prompt and delayed signals is very distinctive and easy to distinguish from background, making a practically-sized antineutrino detector possible to build.

This paper discusses the instrumentation located within the Daya Bay antineutrino detectors. The target mass monitoring systems, discussed in section~\ref{sec:levelinstrumentation}, track changes in the target mass of the ADs in situ and in real time during physics data collection.  An accurate determination of the target mass is critical to the analysis of data from an antineutrino detector, as the expected IBD event rate is proportional to the number of target protons in the detector. Additional instrumentation, discussed in section~\ref{sec:additionalinstrumentation}, monitors the general health and stability of the detectors.

\section{Antineutrino detector design}

\begin{figure}[t]\hfil\subfloat[A cross-section of a filled detector, taken through the mineral oil overflow tanks.]{
\includegraphics[width=74mm]{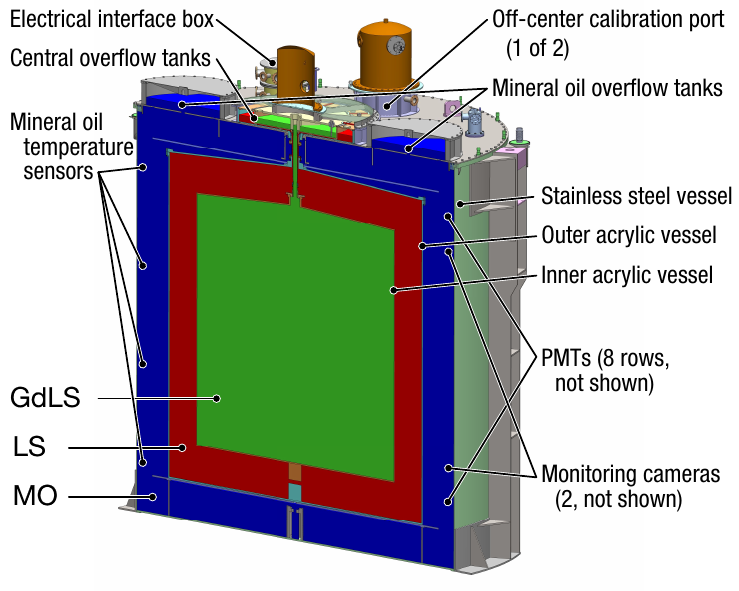}}\hfill
\subfloat[Close-in section of the detector and overflow tanks, taken through the calibration ports.]{\includegraphics[width=74mm]{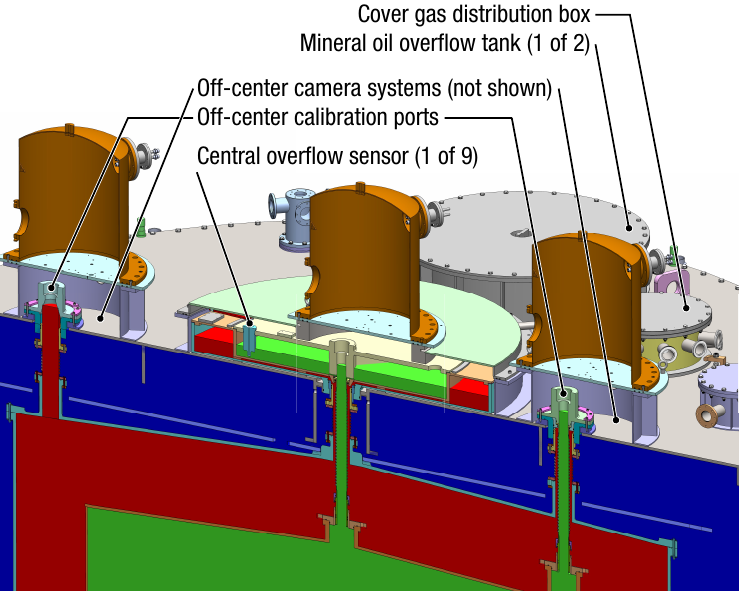}}\hfil%
\caption{Overview drawings of a Daya Bay antineutrino detector, showing the nested liquid volumes and locations of instrumentation and key reference points. The stainless steel vessel of each AD is $5\unit{meters}$ in diameter and $5\unit{meters}$ tall, the outer acrylic vessel is $4\unit{m}$ tall and $4\unit{m}$ in diameter, and the inner acrylic vessel is $3\unit{m}$ tall and $3\unit{m}$ in diameter. Colors distinguish the different liquid regions: gadolinium-doped liquid scintillator (GdLS) is shown in green, plain liquid scintillator (LS) in red, and mineral oil (MO) in blue.}\label{fig:ADoverview}
\end{figure}

The Daya Bay antineutrino detector (AD) design shown in Figure~\ref{fig:ADoverview} has been optimized for detecting antineutrinos via inverse beta decay (IBD) events. It consists of three nested volumes, with two acrylic vessels containing the inner liquids. The innermost volume contains approximately 20 tons of gadolinium-doped organic liquid scintillator (``GdLS''), serving as the IBD target for antineutrino capture. The liquid scintillator is primarily linear alkyl benzene (LAB). Hydrogen atoms in the LAB are the primary targets for antineutrino capture. The middle region contains about $20\unit{tons}$ of unloaded scintillator (the ``LS volume'') acting as a gamma catcher, ensuring that high-energy photons produced in the target volume are reliably converted to scintillation light. The outer volume, filled with about $36\unit{tons}$ of inert mineral oil (MO), provides a buffer between the photomultiplier tubes (PMTs) and the scintillator, improving PMT light collection from the target region and shielding the scintillating liquids from the radioactivity of the PMT materials. Above the main detector volumes are overflow tanks, shown in Figure~\ref{fig:OFcutaway}, which provide space for thermal expansion of the detector liquids. More information on the detector design can be found in \cite{DYBad12}, on the acrylic vessels in \cite{2012arXiv1202.2000B}, and on the scintillator properties and production in~\cite{Yeh2007329,Ding2008238}.

The detector design makes predicting the expected event rate from the antineutrino flux straightforward. The three distinct volumes, physically separated with transparent acrylic walls, eliminate the need for fiducial volume cuts, reducing cut uncertainty and maximizing detection efficiency. The target volume, doped with gadolinium, is the only region that contributes meaningfully to the measured IBD event rate \cite{Guo:2007ug}, and so the number of IBD targets in the target volume directly determines the predicted event rate.\footnote{Geometrical spill-in and spill-out effects, from IBD events occurring at the edge of the GdLS region, are accounted for accurately using Monte Carlo simulations~\cite[\textsection5.5]{DYBad12}. They can affect the event rate by up to 5\%~\cite{DYBcpc}.} Only free protons (hydrogen atoms) are useful IBD targets: while antineutrinos do initiate IBD reactions on protons bound in nuclei, the produced neutron generally is still bound and cannot be easily detected, and the positron and photon signals are difficult to distinguish from accidental backgrounds. The free proton count in the target is directly proportional to the mass of liquid. Chemical analysis of the target liquid measures the carbon-to-hydrogen ratio, which is used together with the target mass to determine the total number of target protons in the detector.

\subsection{Overflow tanks}

\begin{figure}[tb]
\centering
\includegraphics[clip=true, trim=3.125in 4.125in 3.125in 1.875in, width=\textwidth]{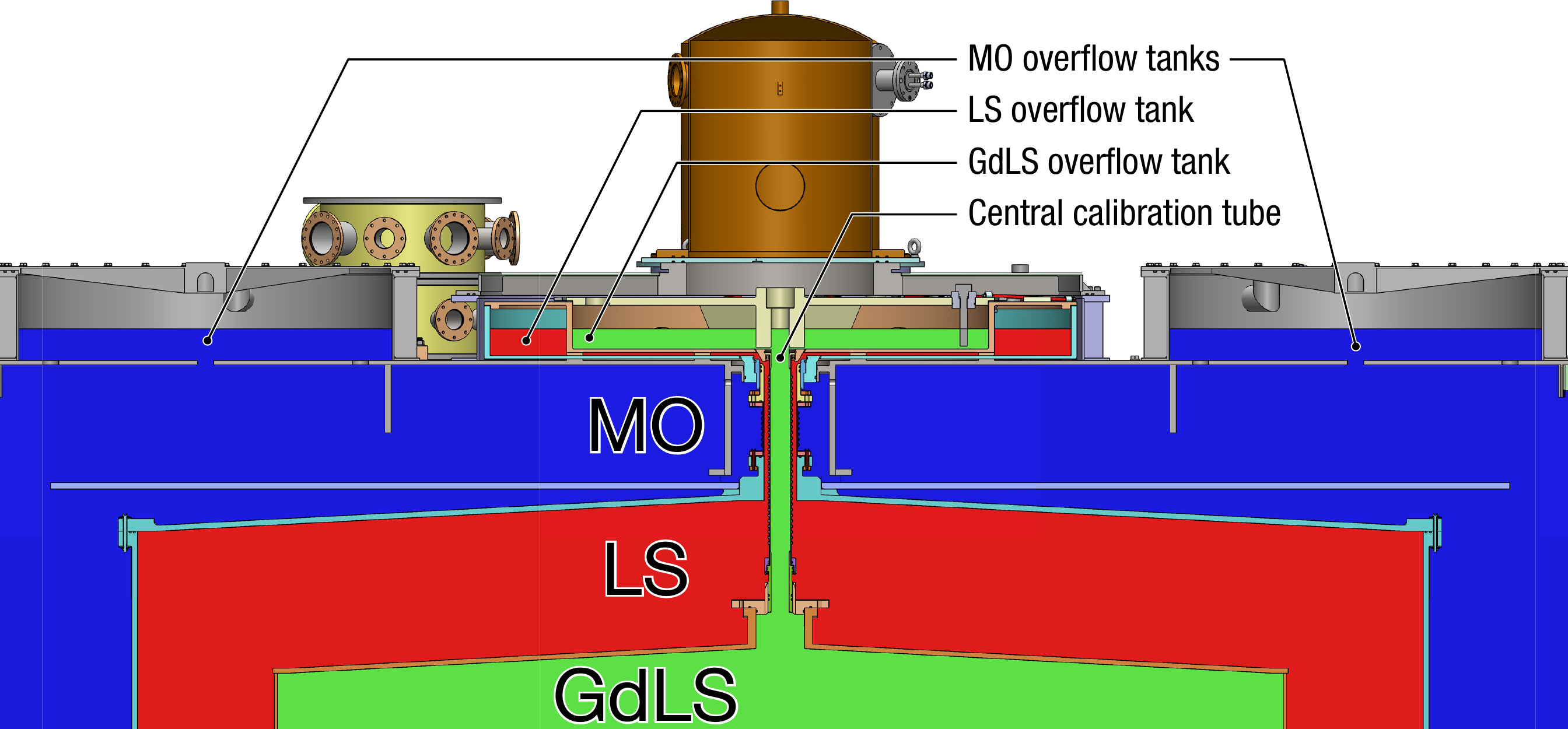} 
\caption{A close-up view of the central and mineral oil overflow tanks, showing the two nested central overflow tanks and the liquid communication between the overflow tanks and the main detector volume. Mineral oil enters its overflow tanks through holes in the vessel lid. GdLS enters the inner central overflow tank through a bellows that passes through the middle of the central calibration port connecting the inner volumes to the central overflow tank. LS occupies the exterior portion of the calibration port tube, connecting to the outer central overflow tank.}\label{fig:OFcutaway}
\end{figure}

The Daya Bay ADs are subjected to variations in temperature and external pressure during their construction and deployment. A completely sealed detector would be unable to tolerate any changes in external conditions; as the detector liquids expanded or contracted due to thermal changes, the resulting pressure differentials and stresses in the acrylic vessels could damage or destroy the detector. To avoid this problem, each detector volume has one or more overflow tanks located on top of the detector, as shown in Figures~\ref{fig:ADoverview},~\ref{fig:OFcutaway}, and~\ref{fig:ADlidlayout}. Each overflow tank is  directly connected with the main volume for that liquid type, as shown in Figure~\ref{fig:OFcutaway}; during detector filling the liquids are filled until the overflow tanks are partly full\citefillingpaper. The excess fluid capacity provided by the overflow tanks ensures that the main detector volumes are always completely full of liquid.
The central overflow tank contains two separated nested acrylic tanks. The inner region, the GdLS overflow tank, is $1285\unit{mm}$ in diameter and
$155\unit{mm}$ deep. It is surrounded by the LS overflow tank, $1790\unit{mm}$  in diameter and $145\unit{mm}$ deep. These dimensions give maximum capacities of approximately $200$ {liters} of GdLS and $165$ {liters} of LS.
The mineral oil overflow tanks are each $1150\unit{mm}$ in diameter and $250\unit{mm}$ tall, giving a capacity of $260\unit{liters}$ of MO per tank or $520\unit{liters}$ per detector.
A cover gas system supplies a continuous flow of dry nitrogen gas to the empty volume of each overflow tank, maintaining a stable, oxygen-free environment for the liquid scintillator \cite{covergaspaper}. 

The range of temperatures that the overflow tanks can buffer against is limited. During detector filling, the overflow volumes are filled to approximately one third of maximum capacity, giving an operational temperature range of $23^{+4}_{-2}\,\degC$ for the filled detectors\citefillingpaper{}\footnote{Based on the thermal expansion coefficient of GdLS, measured as $\Delta V/V = (9.02\pm0.16)\times10^{-4}{\unit{K}}^{-1}$ at $19\degC$ \cite{minfangemail}.}. The typical operating temperatures of 22.5--23.0\degC{} have been well within this range\footnote{Temperature variations between the different water pools are larger than the variations of a single pool; the temperature of each  pool is typically controlled to $\pm0.2\degC$ or better.} (see also Figure~\ref{fig:monitoringdata}). The instrumentation described in this paper continuously monitors the levels and temperatures of the fluids in all overflow tanks, ensuring that conditions remain within the operational range of the detectors at all times.

\section{Stability and monitoring requirements}\label{sec:requirements}

The design goals of the Daya Bay experiment \cite[\textsection3.2]{Guo:2007ug} specify a baseline target mass uncertainty of 0.2\% ($2000 \unit{ppm}$) and goal target mass uncertainty of 0.02\% ($200 \unit{ppm}$) on a nominal target mass of $\text{20,000}\unit{kg}$. This requirement corresponds to a baseline uncertainty of $40\unit{kg}$ in the nominal target mass and goal uncertainty of $4\unit{kg}$. The detector filling system\citefillingpaper{} was designed to surpass the goal uncertainty, and the target mass monitoring system described in this paper must be able to match this accuracy goal in order to track longer-term detector changes. Given the central overflow tank diameter of $1285\unit{mm}$ and GdLS density of approximately $860\unit{g/L}$, a $4\unit{kg}$ uncertainty on overflow volume mass corresponds to about a $3.5\unit{mm}$ uncertainty on the overflow tank liquid level. Sensors were chosen with specified uncertainties significantly smaller than this level, minimizing the contribution to the total target mass uncertainty from the overflow tank monitoring.

In practice, after sensor calibrations and accounting for overflow tank geometry uncertainties and overall detector tilt uncertainties, we achieved a total overflow tank mass uncertainty of $2.2\unit{kg}$, corresponding to 0.011\% of the nominal target mass ($110\unit{ppm}$).

\section{Liquid level monitoring instrumentation}\label{sec:levelinstrumentation}

\begin{figure}[!p]%
\def\subfigheight{41mm}%
\hfil%
\subfloat[Top view of the central overflow tanks, showing the mounting locations of the sensors and the support structure of the tanks. The tank's outer diameter is $1.8\unit{m}$.]{\label{fig:ADlidlayoutdiagram}\includegraphics[width=74mm]{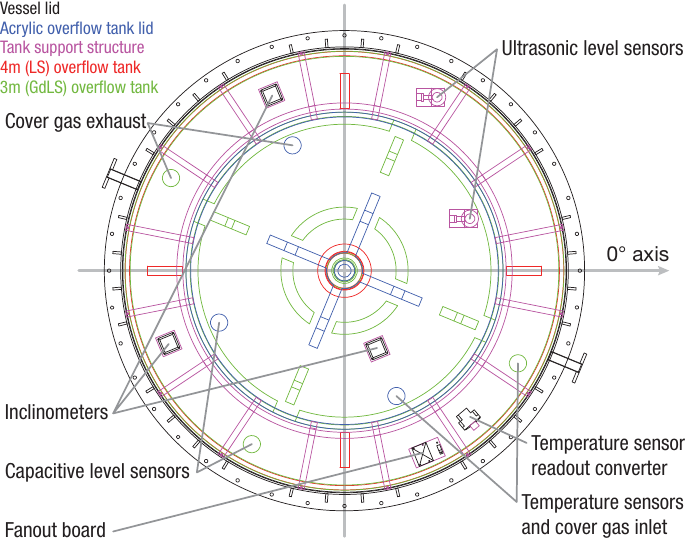}}\hfill
\subfloat[A central overflow tank lid shown at the end of assembly, with all instrumentation installed. (Note that the detector $0\deg$ axis faces the camera in this image.)]{\label{fig:centralOFtankpicture}\includegraphics[width=74mm]{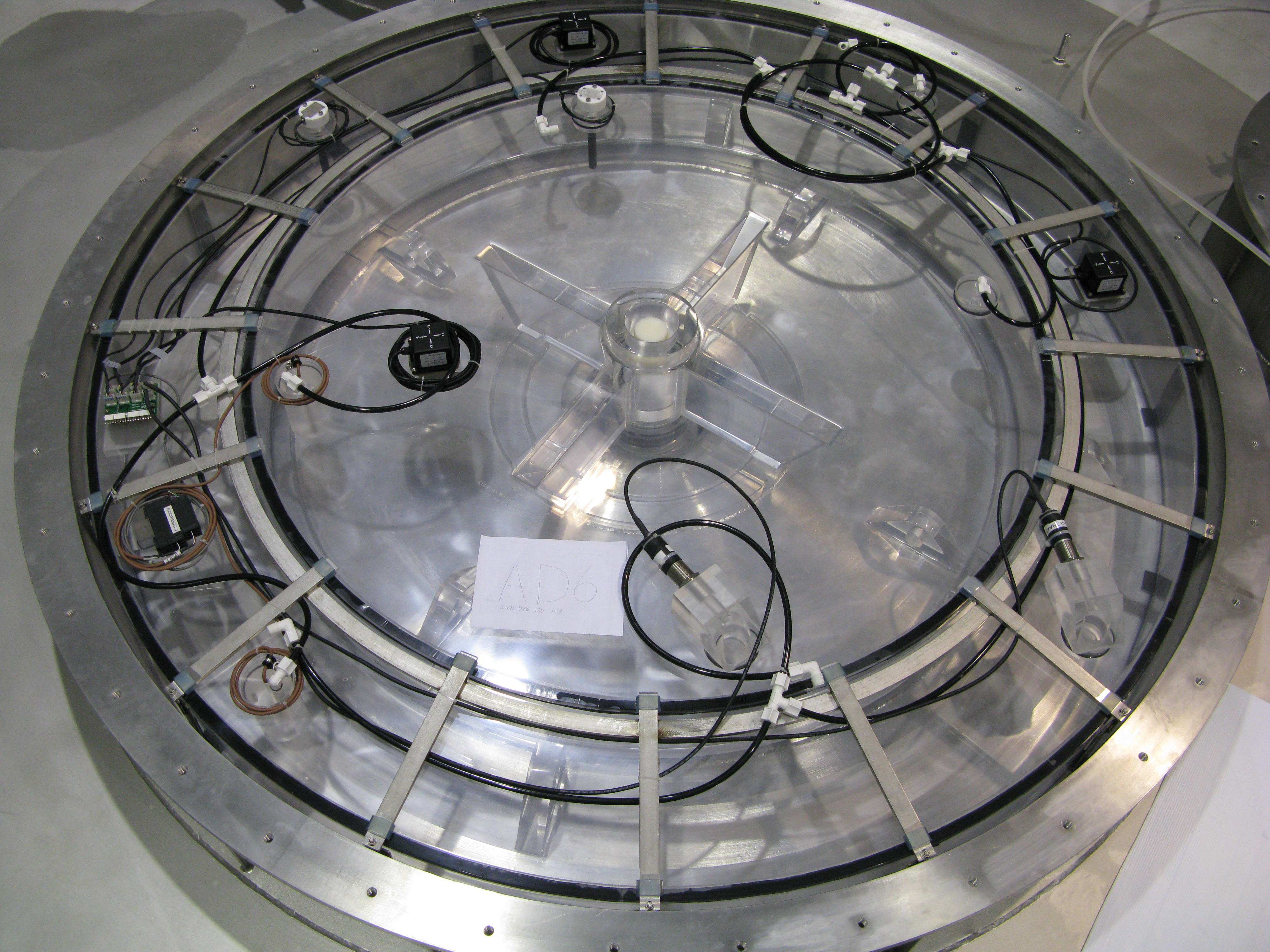}}\hfil%
\\\hfil%
{\captionsetup{justification=raggedright}%
\subfloat[Ultrasonic liquid level sensor.]{\includegraphics[height=\subfigheight]{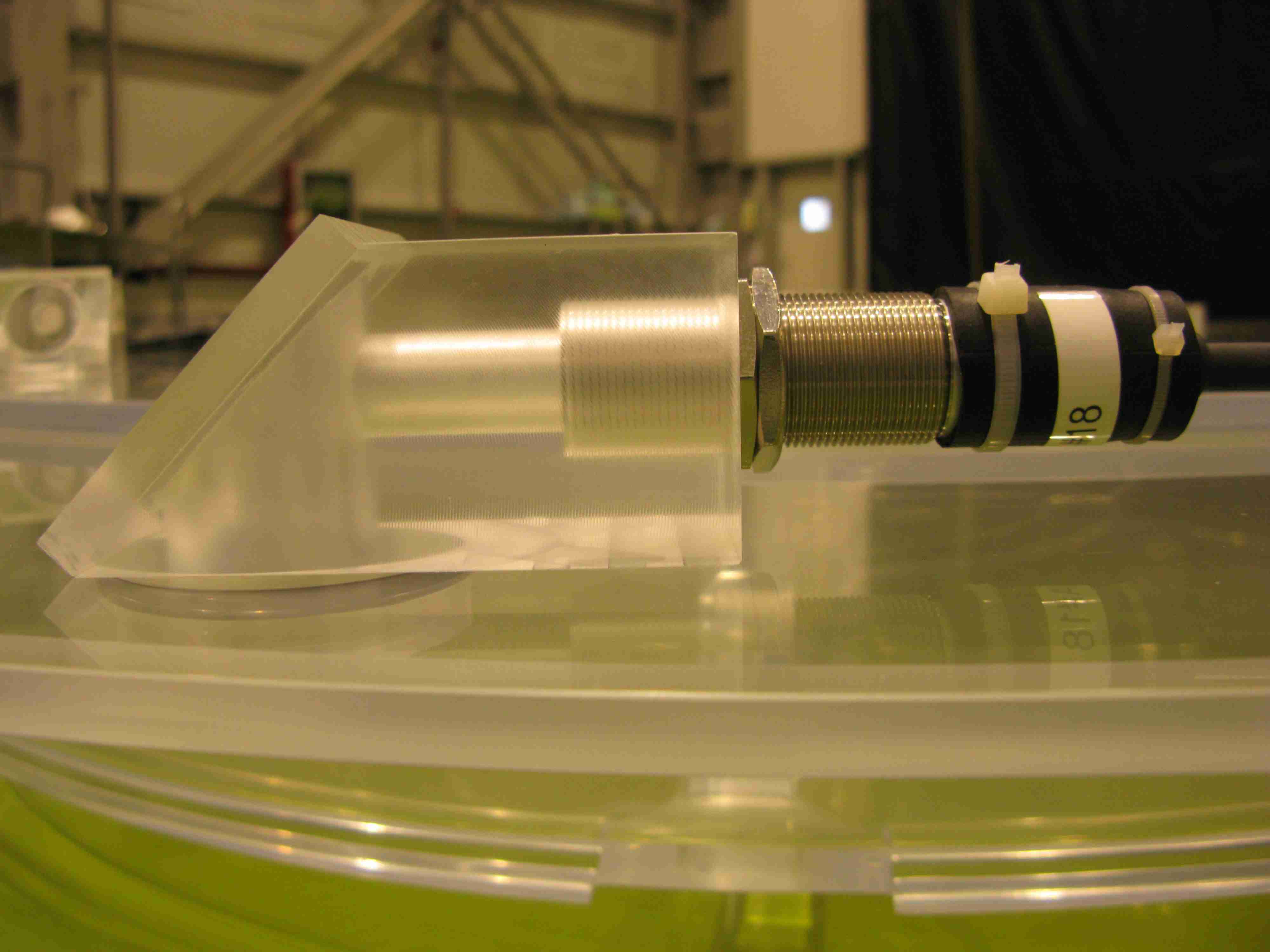}}%
\hfill\subfloat[Temperature sensor.]{\includegraphics[height=\subfigheight,clip=true, trim=20in 10in 17in 0in]{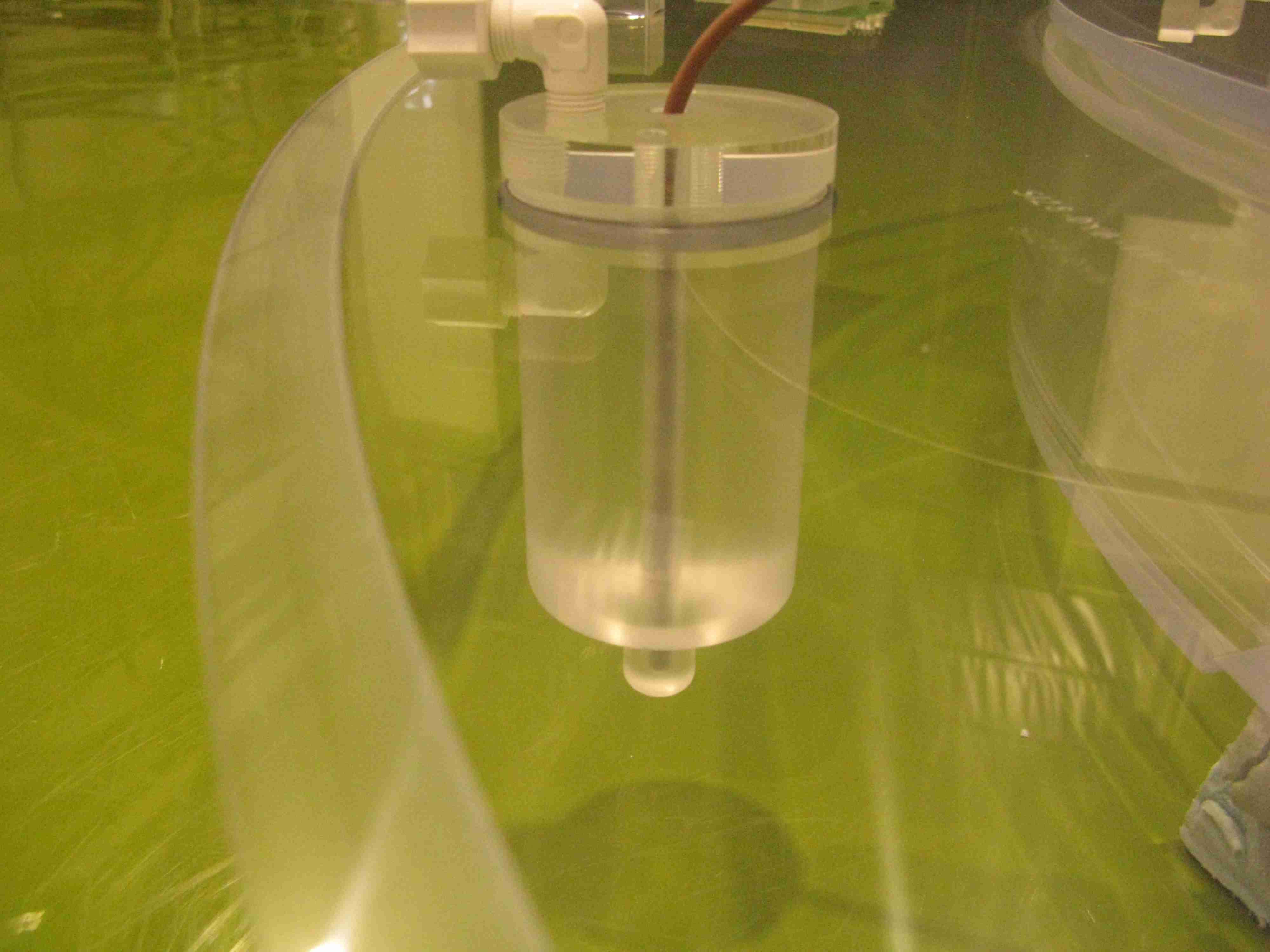}}%
\hfill\subfloat[PTFE capacitive liquid level sensor.]{\includegraphics[height=\subfigheight,clip=true, trim=16in 4in 15in 3in]{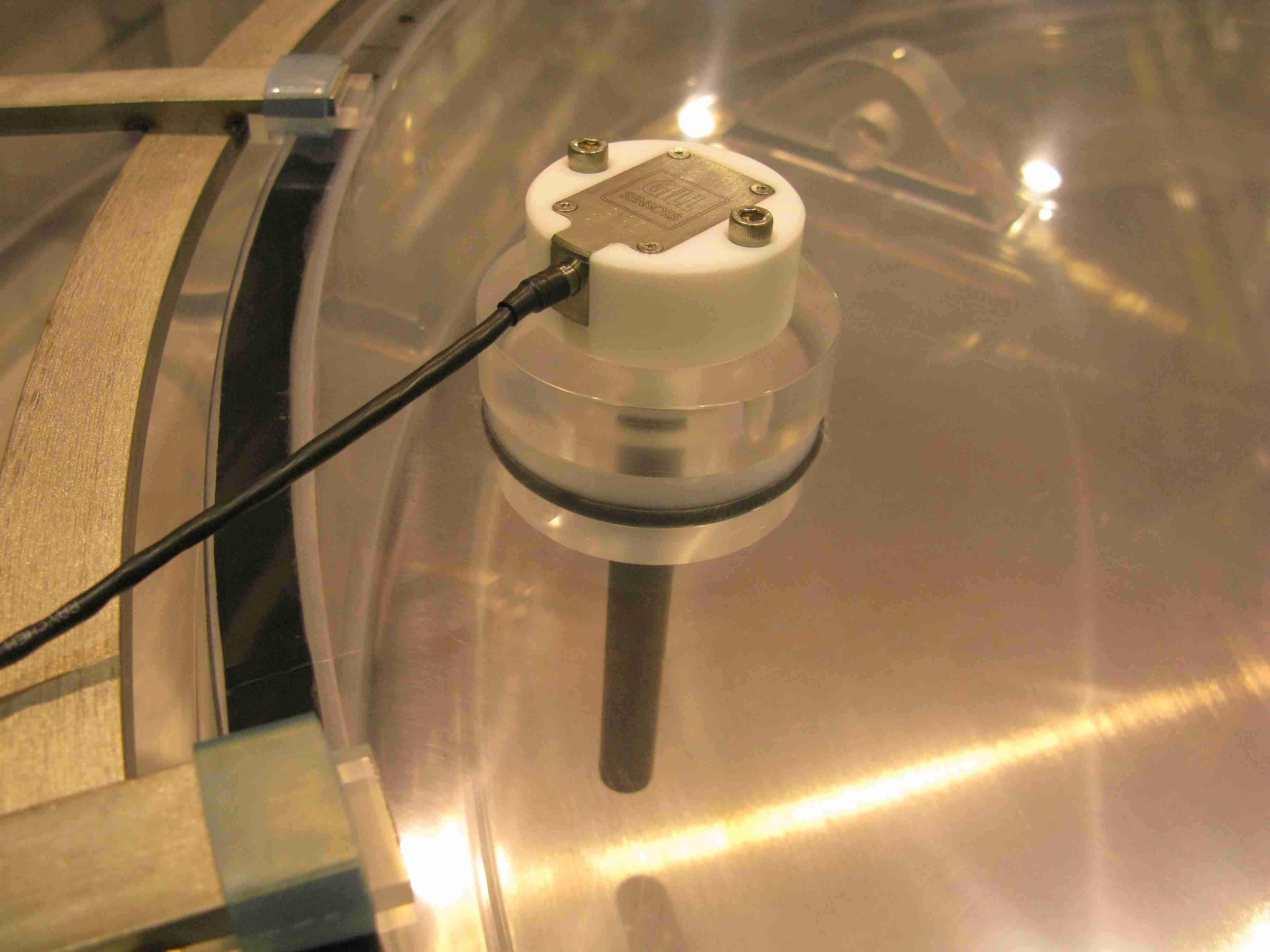}}}%
\hfill\subfloat[Two-axis inclinometer.]{\includegraphics[height=\subfigheight, clip=true, trim=18in 6in 10in 6in]{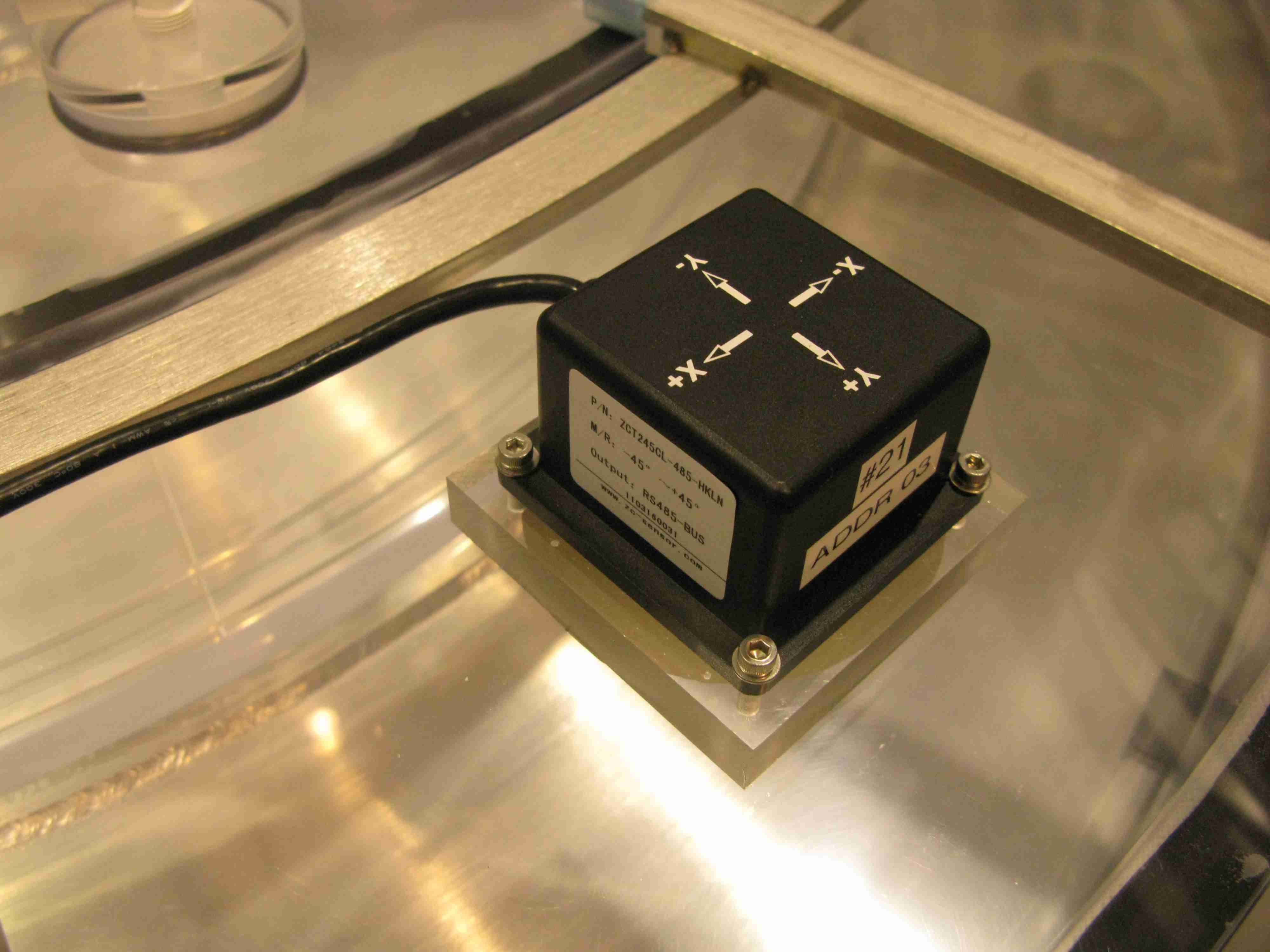}\hfil%
}\\%
\subfloat[A fully instrumented mineral oil overflow tank. The tank's outer diameter is $1.2\unit{m}$.]{\label{fig:MOoverflowtank}\includegraphics[width=\textwidth]{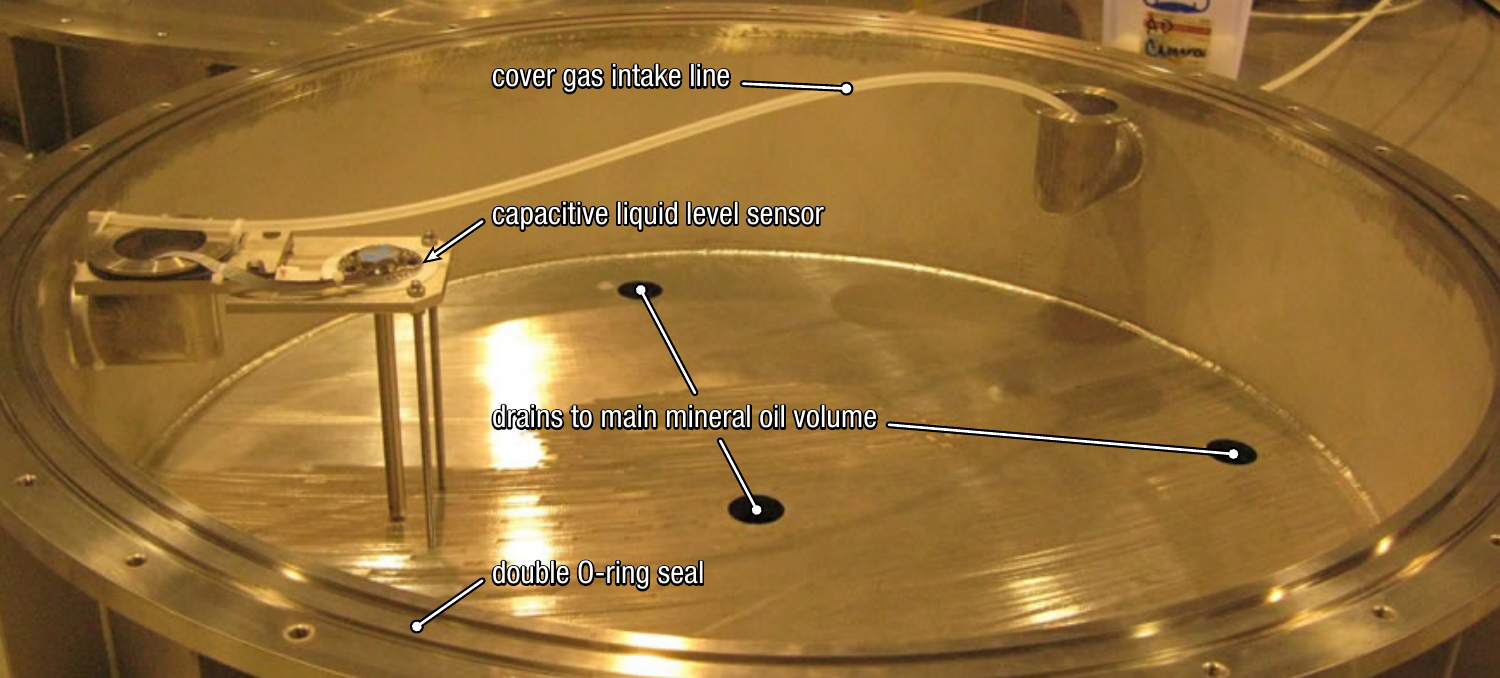}}%
\caption{Instrumentation and layout of the overflow tanks. Top row: layout of the central overflow tank, containing GdLS and LS. Center row: central overflow tank sensors. Bottom: mineral oil overflow tank.}\label{fig:ADlidlayout}
\end{figure}

Three separate systems monitor the liquid height in the detector overflow tanks, as pictured in Figure~\ref{fig:ADlidlayout}. The two central overflow tanks are each instrumented with an ultrasonic liquid level sensor and a capacitive liquid level sensor. In addition, cameras in the off-center calibration ports monitor the GdLS and LS liquid levels. One of the two mineral oil overflow tanks contains a capacitive liquid level sensor.  The GdLS and LS overflow tanks each have two independent sensors measuring the current liquid level to provide redundancy in the event of a sensor failure. The mineral oil level is less critical and is only monitored by a single sensor. The monitoring cameras provide a cross-check against potential long-term drift in the GdLS and LS sensors.

\subsection{Ultrasonic level sensors}

\begin{figure}[tb]\centering
\includegraphics[height=65mm,clip=true,trim=0 6in 0 6in]{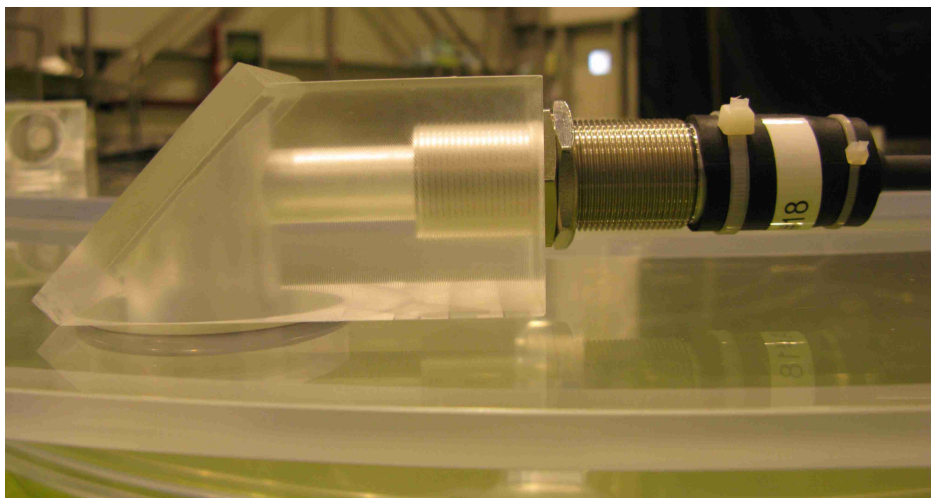}
\caption{An ultrasonic distance sensor shown mounted with rear light baffle on an overflow tank during detector assembly. The acrylic reflector is visible at left. (See also Figure~\protect\ref{fig:ultracalibrationstand}.)}\label{fig:ultrasonic}
\end{figure}

The ultrasonic liquid level sensors, which measure the distance from the sensor to the target by timing the echo of an ultrasonic pulse, are commercial Senix Corporation ToughSonic TSPC--30S1 distance sensors \cite{senixcorp}. They are non-contact sensors, which is ideal for minimizing disruptions to the scintillator and alleviating materials compatibility concerns. As the fluid level in the overflow tank rises, the distance measured by the ultrasonic sensor decreases linearly, proportional to the increase in fluid height. Two sensors are located in each central overflow tank, one in the GdLS volume and one in the LS volume. As shown in Figure~\ref{fig:ultrasonic}, they are mounted horizontally on the top of the overflow tank lid, with the ultrasound beam directed downwards by a flat acrylic reflector located in the mount. This arrangement was necessitated by space limitations; we found that the addition of the reflector caused no discernible change in the performance of the sensor. Figure~\ref{fig:ultrasonic} also shows a light baffle on the back of the sensor, needed to block light from the sensor's rear status LEDs from reaching the detector PMTs.

As described in section~\ref{sec:ultrasoniccalibration}, each ultrasonic sensor was individually calibrated. After calibration, the ultrasonic sensors are the most accurate liquid level sensors, and are our primary liquid height reference.

\subsection{Capacitive level sensors}

\begin{figure}[t]\hfil
\subfloat[A stainless steel capacitive liquid level sensor, located in a mineral oil overflow tank.]{\label{fig:MOcapsensor}%
\includegraphics[clip=true, trim=4.25in 1in 0.8125in 2in, height=70mm]{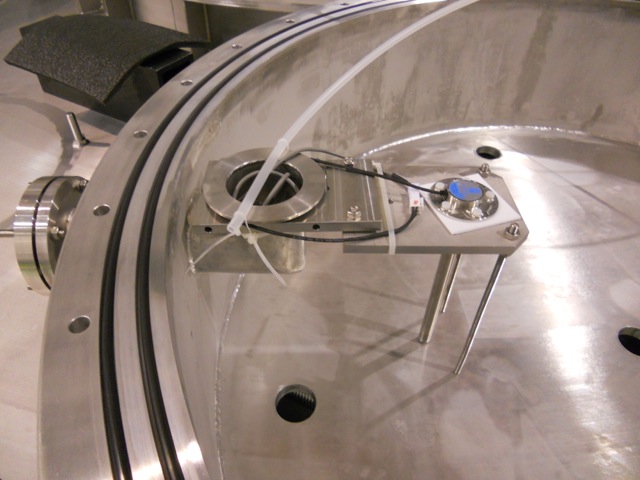}%
}\hfill
\subfloat[A PTFE capacitive liquid level sensor, located in the central overflow tank.]{\label{fig:tefloncapsensor}%
\includegraphics[clip=true, trim=6in 0in 6in 0in, height=70mm]{figures/central_cap_mounted.jpg}}\hfil%
\caption{The two styles of capacitive liquid level sensor, shown mounted during detector assembly.}\label{fig:capsensors}
\end{figure}

The three capacitive liquid level sensors, provided by Gill Sensors \cite{gillsensorsco}, are customized versions of automotive fuel tank sensors. They consist of two concentric cylinders mounted vertically in the liquid tanks as shown in Figures~\ref{fig:ADoverview} and~\ref{fig:capsensors}. The volume between the cylinders is open to the liquid, and as the tank level rises this area fills in with liquid. The dielectric properties of the liquid cause a change in the capacitance between the two cylinders, which is read out by instrumentation in the top of the sensor. The sensor outputs a value in counts, representing the fraction of the sensor that is immersed in liquid, with 0 as the minimum value. Nominally a 10-bit ADC is used, giving a resolution of 1024 counts over the full range of the sensor. In practice, we observed sensor readings of 1050 counts or more, suggesting a more complicated signal processing scheme inside the sensor. Each sensor was individually calibrated as described in section~\ref{sec:capcalibration}.

There are two types of capacitive sensors in each detector. One mineral oil overflow tank per detector contains a single commercially available stainless steel R-Series sensor, shown in Figure~\ref{fig:MOcapsensor}. Its only customization is a factory calibration to the dielectric properties of mineral oil. The R-Series sensors are $16\unit{mm}$ in diameter and $235\unit{mm}$ long; $220\unit{mm}$ of their length is sensitive to liquid level changes. Stainless steel is not suitable for contact with gadolinium-doped scintillator, so the GdLS and LS volumes in the central overflow tank use custom M-Series sensors from Gill, shown in Figure~\ref{fig:tefloncapsensor}. The only wetted material in these sensors is chemically inert polytetrafluoroethylene (PTFE). Ordinary white PTFE makes up the head of the sensor, enclosing the readout electronics, while conductive carbon-impregnated black PTFE makes up the capacitive body of the sensor. Both varieties of PTFE are chemically compatible with the Daya Bay liquid scintillator over long periods of time. The M-series sensors are $16\unit{mm}$ in diameter and $153\unit{mm}$ long (active region $144\unit{mm}$). They were provided with a factory calibration for linear alkyl benzene.

The factory calibrations of all sensors were supplemented with a laboratory calibration, described in section~\ref{sec:capcalibration}. In practice, the capacitive sensors have coarser resolution and are not as reproducible as the ultrasonic sensors, so they are used as secondary sensors in the event that an ultrasonic sensor fails. They also have the advantage of being insensitive to the gas composition in the overflow volumes, since they do not depend on the speed of sound as the ultrasonic sensors do. Along with the off-center cameras, they provide a cross-check on the ultrasonic sensor data, ensuring the ultrasonic sensor response is not changing over time.

\subsection{Off-center cameras}

\begin{figure}[tb]\hfil%
\subfloat[Model view of the off-center camera system.]{\label{fig:DGUTcameraschematic}\includegraphics[width=70mm]{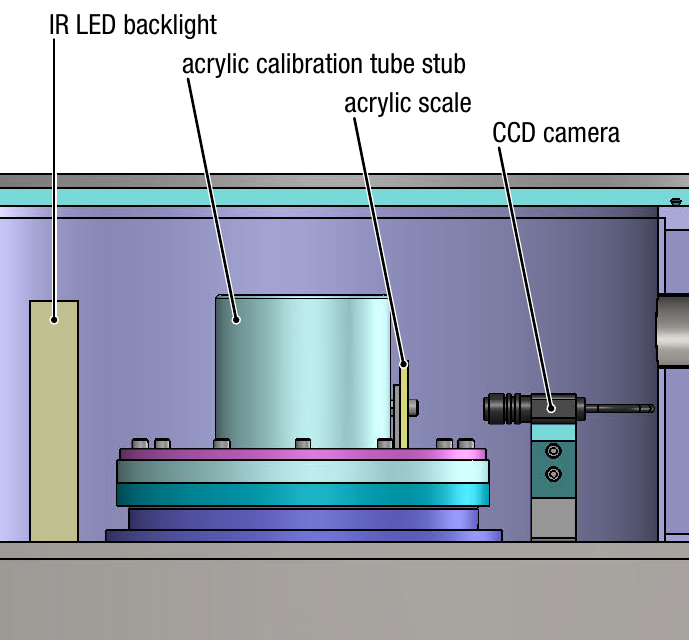}}\hfil%
\subfloat[An assembled off-center camera system in AD8.]{\label{fig:DGUTcameraphoto}\includegraphics[width=70mm]{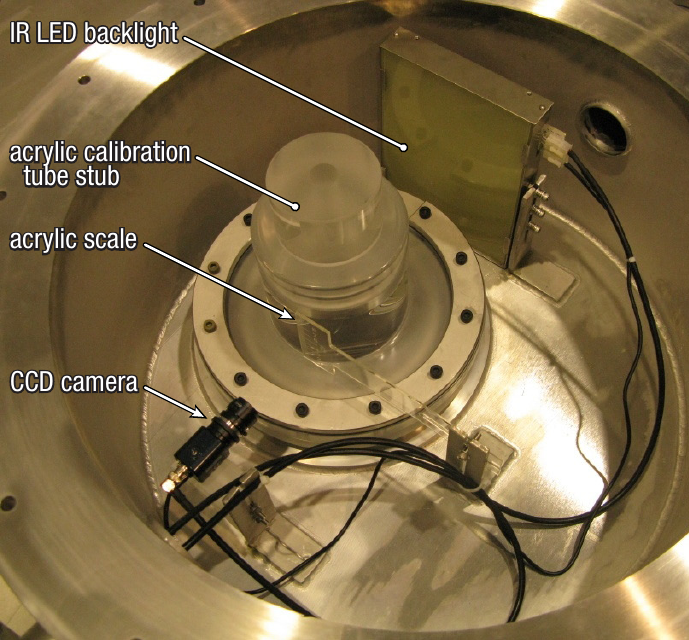}}\hfil%
\caption{Images of the off-center camera systems. There are two cameras in each detector to monitor the GdLS and LS liquid levels at the off-center calibration ports. See also Figures~\protect\ref{fig:ADoverview} and~\protect\ref{fig:DGUTcameraimages}.}\label{fig:DGUTcamerasystem}
\end{figure}

\begin{figure}[tb]\hfil%
\subfloat[Off-center camera image of the GdLS level in AD3. Note that the GdLS calibration tube is straight.]{\label{fig:GdLSimage}\includegraphics[width=70mm]{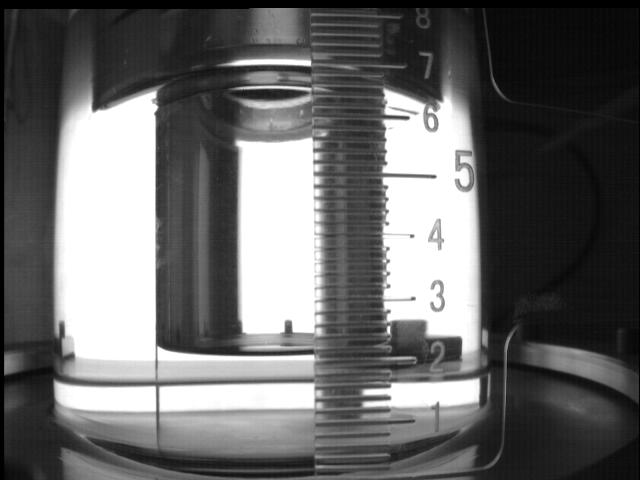}}\hfil%
\subfloat[Off-center camera image of the LS level in AD3. Note that the LS calibration tube has beveled sides.]{\label{fig:LSimage}\includegraphics[width=70mm]{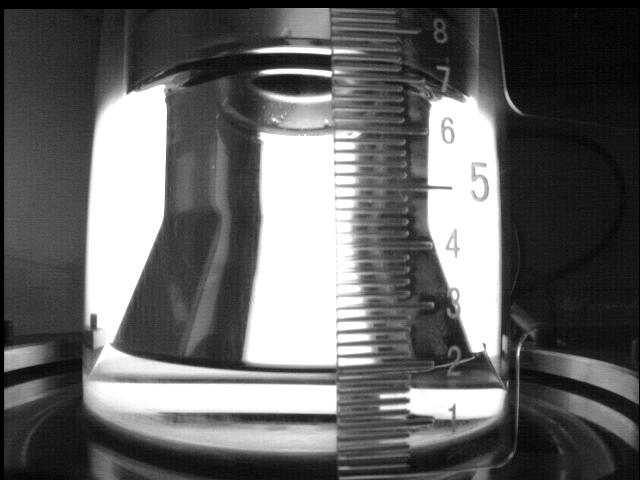}}\hfil%
\caption{Sample images taken by the off-center camera system.}\label{fig:DGUTcameraimages}
\end{figure}

There are two camera systems located on the detector lid to monitor the GdLS and LS liquid levels.
As shown in Figure~\ref{fig:ADoverview}, they are located in the off-axis calibration ports of the detector, and view the liquid level in the calibration tubes.
Figure~\ref{fig:DGUTcamerasystem} shows an overview and picture of the system.
Sample images from the cameras are shown in Figure~\ref{fig:DGUTcameraimages}.
A ruled scale placed between the cameras and the calibration tubes provides a reference scale for tracking fluid level changes.
This enables the cameras to be used as a cross-check against drifts of the other sensors.
The position of the camera elements and calibration tubes was surveyed during installation as part of the global detector survey, so the relative positions of the camera and visible sections of the calibration tubes are known to millimeter accuracy.

Infrared-sensitive cameras were chosen with the goal of being able to run the level monitoring system during detector operation, without interference with the detector PMTs.
The cameras are model Watec WAT-902H2 Supreme, equipped with a 380-kilopixel  monochromatic infrared-sensitive CCD sensor.
Illumination for each camera is provided by an array of infrared LEDs behind a diffusing screen, facing the camera from behind the calibration tube.
The array produces $30\unit{mW}$ of infrared illumination at the camera using $300\unit{mW}$ of electrical power.
The LED spectrum is peaked at $886\unit{nm}$ with some emittance over the range of $800$ to $1000\unit{nm}$.
This spectrum overlaps with the camera CCD's near-infrared sensitivity, but is beyond the detector PMT sensitivity (Hamamatsu R5912 PMT response falls off at $700\unit{nm}$) and the absorption spectrum of the scintillator, which does not extend significantly beyond $400\unit{nm}$. 
While dark box tests confirmed that the infrared LEDs did not trigger the PMTs, in-situ testing revealed enough of an increase in PMT hit rates (up to $30\unit{Hz}$) that as a precaution the cameras are not operated while the PMTs are in use. 
The off-axis camera systems are thus only operated to cross-check the other liquid sensors when physics data is not being collected.

\section{Additional instrumentation}\label{sec:additionalinstrumentation}

\subsection{Inclinometers}

Three inclinometers are mounted on the detector lid, in positions shown in Figure~\ref{fig:ADlidlayoutdiagram}. Each sensor, a Shanghai Zhichuan Electronic Tech Company Ltd.\ model ZCT245CL-485-HKLN, has a nominal accuracy of $\pm0.05\deg$ with a resolution of $0.01\deg$ \cite{inclmanual}. An inclinometer is pictured in Figure~\ref{fig:inclinometerphoto}. The sensors measure the absolute tilt of the detector from the local vertical. Each sensor measures the tilt of two axes: the deviation of both its local $x$-axis and its local $y$-axis from the plane perpendicular to the local vertical. All three sensors are mounted with the same $x$- and $y$-axis directions to facilitate cross-checking between sensors. Three sensors are used so that both global tilts of the entire detector and localized lid deformations, such as flexing or deflection, can be resolved. To first order, the calculation of the liquid level in the overflow volume assumes the liquid is level; the inclinometer data allows us to quantify any deviation from levelness and apply a correction if necessary.

\subsection{Overflow temperature sensors}\label{sec:OFtempsensors}

\begin{figure}[b]\hfil
\subfloat[A close-up image of two Pt100 temperature sensors. Each sensor is $6.35\unit{mm}$ in diameter.]{\label{fig:Pt100}\includegraphics[clip=true, trim=4in 14.5in 29in 10in, height=54mm]{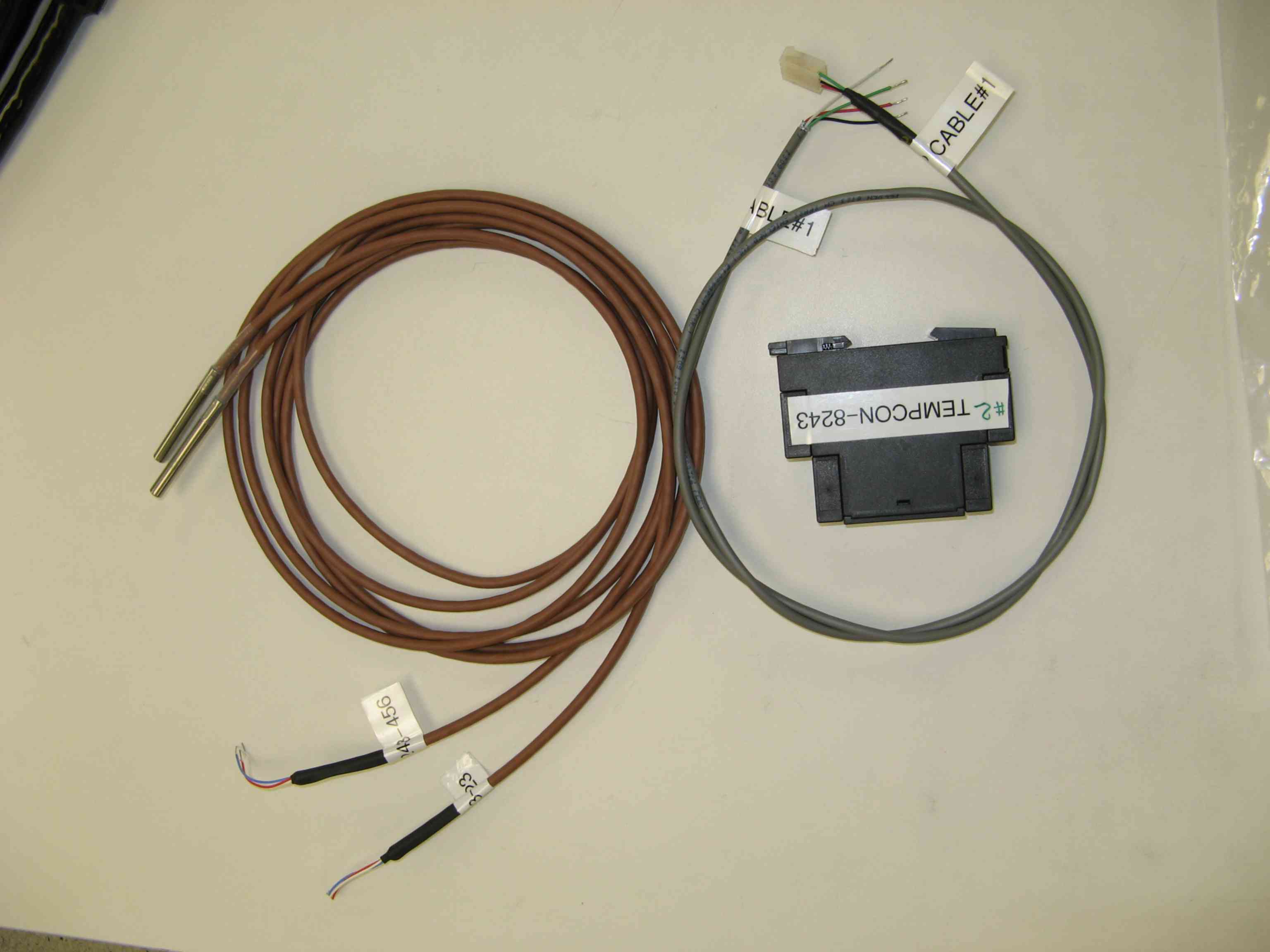}}\hfil
\subfloat[A temperature sensor installed in its acrylic mount, shown during overflow tank assembly. The elbow at upper left is the cover gas intake connection.]{\label{fig:Pt100mount}\includegraphics[clip=true, trim= 8in 10in 8in 0in, height=54mm]{figures/IMG_1676.jpg}}\hfil%
\caption{The Pt100 temperature sensors (left) and mounted in place (right). A stainless steel case encloses each sensor, and a PTFE sleeve protects the junction between the casing and the sensor cable. }\label{fig:tempsensors}
\end{figure}

The central overflow volumes for both GdLS and LS each contain a single type Pt100 100-{\ohm} platinum resistance thermometer. The sensors, shown in Figure~\ref{fig:Pt100}, were provided by HW Group and are DIN class A (accuracy specification $\pm 0.2 \degC$ at $25\degC$). An acrylic mounting well, shown in Figure~\ref{fig:Pt100mount}, holds each sensor in position and prevents its stainless steel casing from contacting the detector liquids. A drop of mineral oil placed in the well creates a good thermal connection between the sensor and the acrylic.

Temperature sensor readout is performed using a HW Group model Temp-485-2xPt100  readout interface \cite{HWgPt100}. Each detector contains one dual Pt100 readout module mounted on the AD lid that reads out both the GdLS and LS overflow tank temperature sensors. The readout module measures the resistance of the Pt100 sensor and converts it to a digitized temperature reading with a resolution  of $0.01 \degC$, much smaller than the quoted overall accuracy of the sensor, ensuring the absence of quantization effects. Three-wire connections between the sensor and readout module are employed to reduce measurement error from the resistance of the wire leads on the sensor.

\subsection{Internal temperature sensors}\label{sec:MOtempsensors}

\begin{figure}[tb]\hfil
\subfloat[A mineral oil temperature sensor, mounted on the PMT support ladder, with the sensor tip exposed at bottom.]{\includegraphics[clip=true, trim=0in 6in 24in 18in, width=70mm,angle=90]{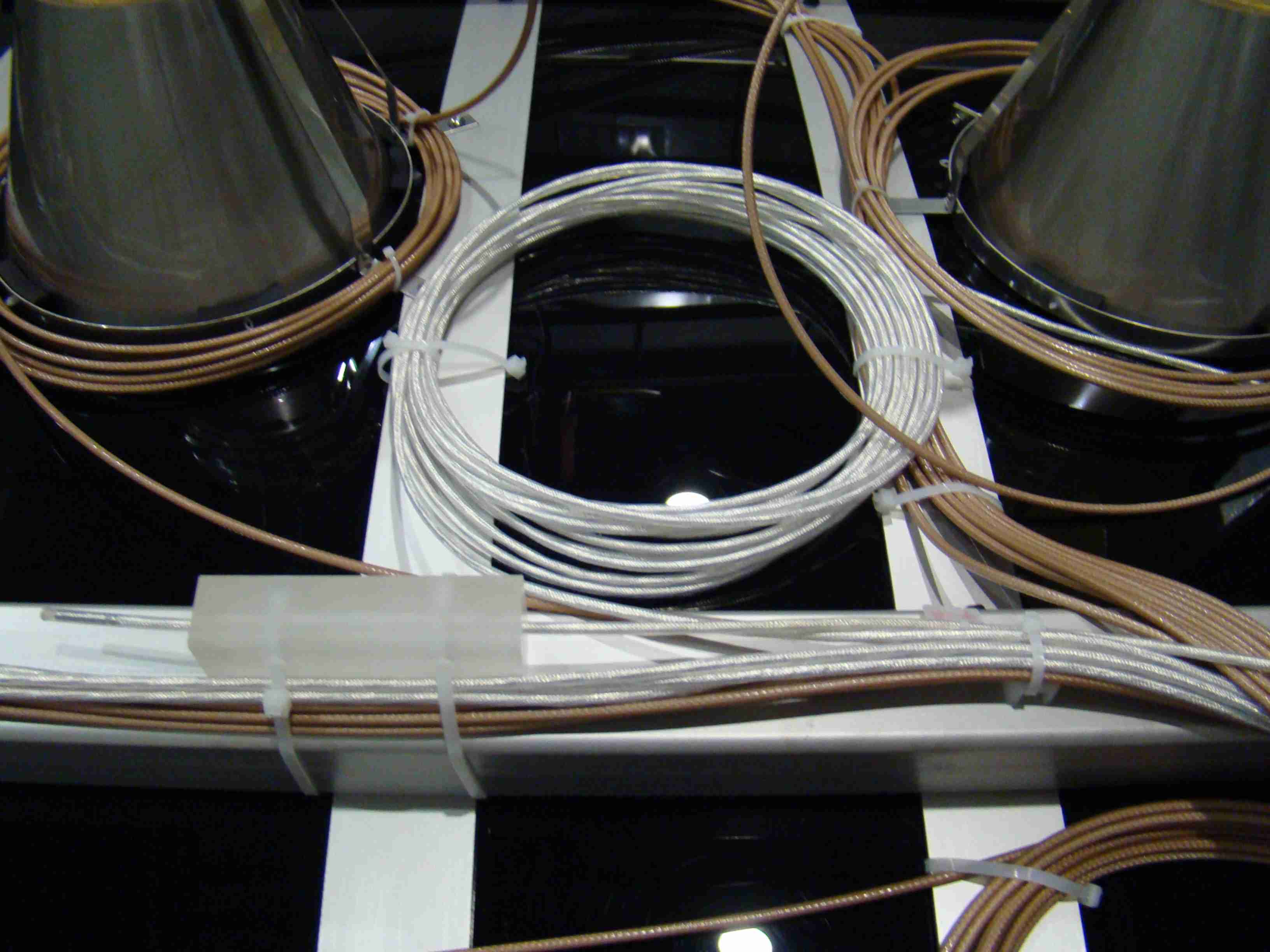}}\hfil
\subfloat[Overview image of an assembled PMT support ladder. Three of the four mineral oil temperature sensors are visible; the fourth is located just beyond the bottom edge of the image.]{\includegraphics[height=70mm]{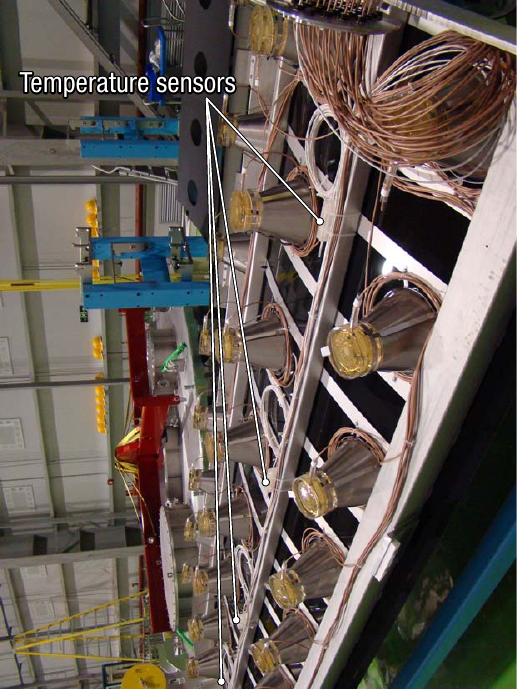}}\hfil%
\subfloat[Model view of the detector, showing the back of the PMT ladders enclosed within the stainless steel vessel. The four mineral oil temperature sensors, circled in yellow, are visible at center right.]{\includegraphics[height=70mm,clip=true,trim=0.5in 0.25in 4in 0.25in]{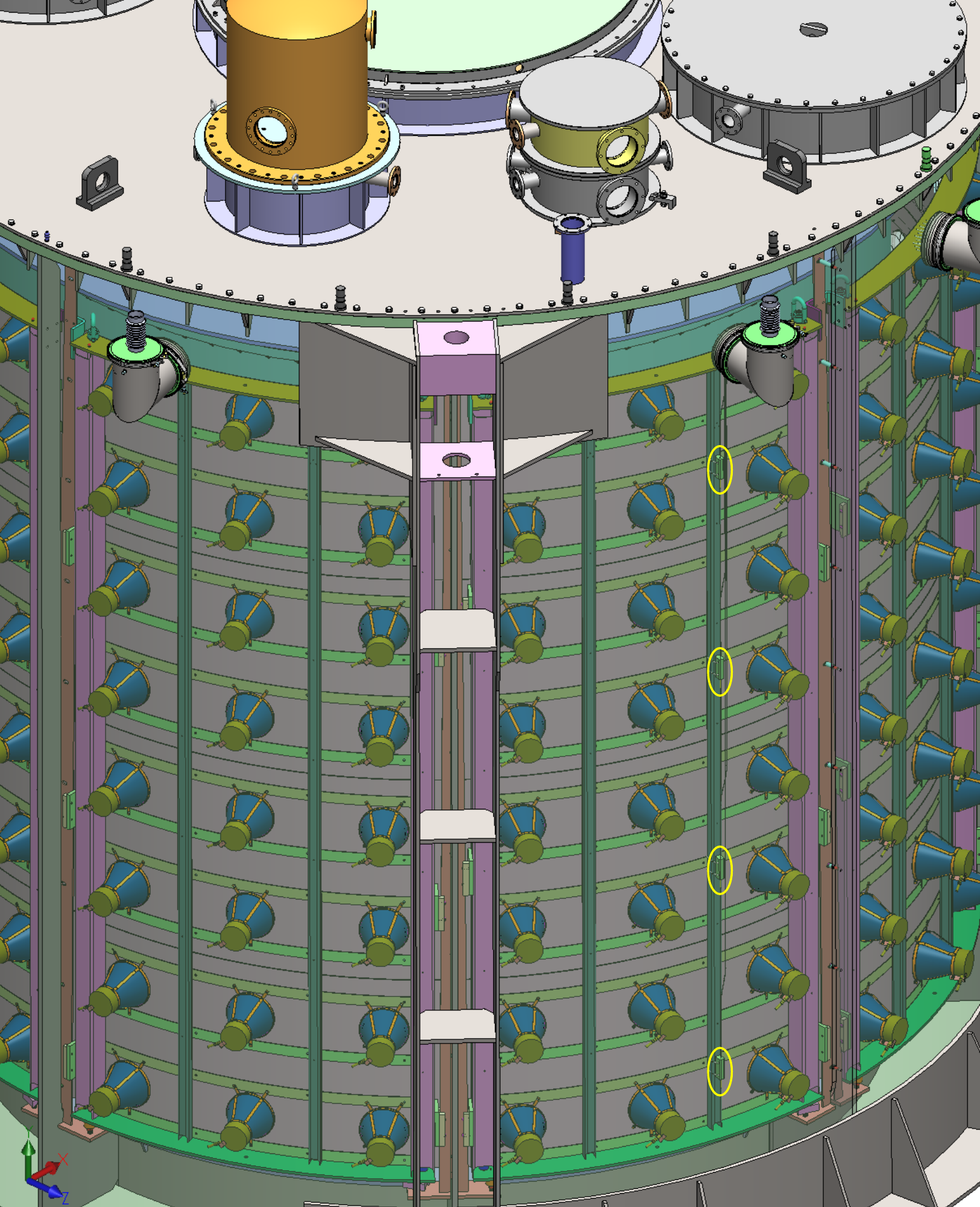}}\hfil%
\caption{Images and mounting locations of the mineral oil internal temperature sensors.}\label{fig:motempsensors}
\end{figure}

The detector contains four temperature sensors mounted in the mineral oil volume, as shown in Figure~\ref{fig:motempsensors}. They are arranged vertically along one PMT support ladder to measure any temperature gradients in the detector resulting from temperature differentials during detector filling or operation. They are electrically and mechanically isolated from the ladder by acrylic housings to give a direct reading of the mineral oil temperature at that location. Each sensor is a Pt100 platinum resistance thermometer, as employed in the overflow tanks. They are connected using a standard four-wire connection to a custom-designed readout module that can accurately measure the sensor's resistance and resolve small changes in temperature.

\subsection{Internal cameras}

Two internal cameras are located on the outer wall of the detector, as shown in Figure~\ref{fig:ADoverview}. One is mounted looking down at the bottom of the target volume, and one is mounted looking up at the top of the target volume. Each camera has a remotely controlled LED bank containing both white and infrared LEDs to illuminate the detector during camera operation. The internal cameras are used extensively during
detector filling and then infrequently afterwards for detector inspection.
They may also be utilized during a future full-volume manual detector
calibration. They cannot be used continually during operation, due to interference between the LED light source and the detector PMTs. Their construction and operation is described in more detail in \cite{camerapaper}.

\section{Sensor calibration}

Each type of sensor measures a different quantity. Some sensors directly measure the quantity of interest, while others return a value that must be converted to physically meaningful units: millimeters of liquid height for the liquid sensors, degrees Celsius for the temperature sensors, or degrees of inclination for the tilt sensors. For many sensors, the conversion between the output quantity and useful units was provided by the manufacturer, but in order to meet the precision level  required in the Daya Bay detectors, we performed laboratory calibrations of all the critical sensors.

\subsection{Ultrasonic liquid level sensors}\label{sec:ultrasoniccalibration}

\begin{figure}[bt]\hfil
\subfloat[Diagram of the ultrasonic sensor calibration stand. The dotted lines show two nominal ultrasound paths. The height $h$ is adjustable from 0 to $\mathbin{\sim}400\unit{mm}$.]{\label{fig:ultracalibrationstand}\includegraphics[width=60mm,height=60mm]{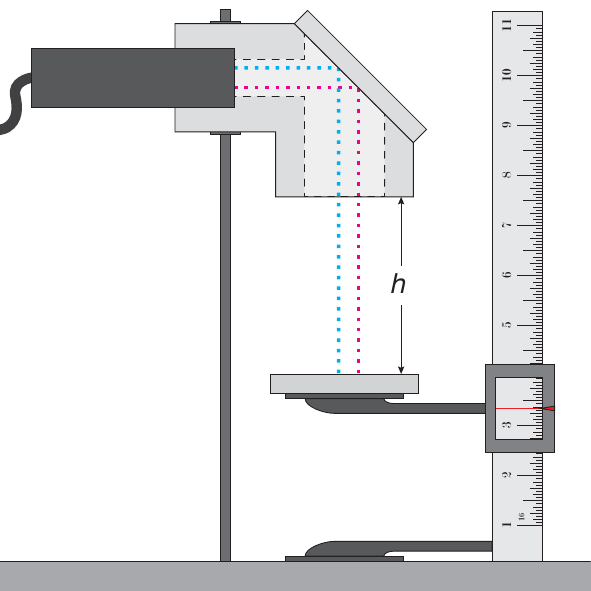}}\hfil
\subfloat[Plot of the calibration data for ultrasonic sensor 1. The data was fit only in the operating region, $h$ from 0 to $150\unit{mm}$. Error bars are the standard deviations of multiple readings taken at each height
.]{\label{fig:ultracalibrationgraph}\includegraphics[height=60mm]{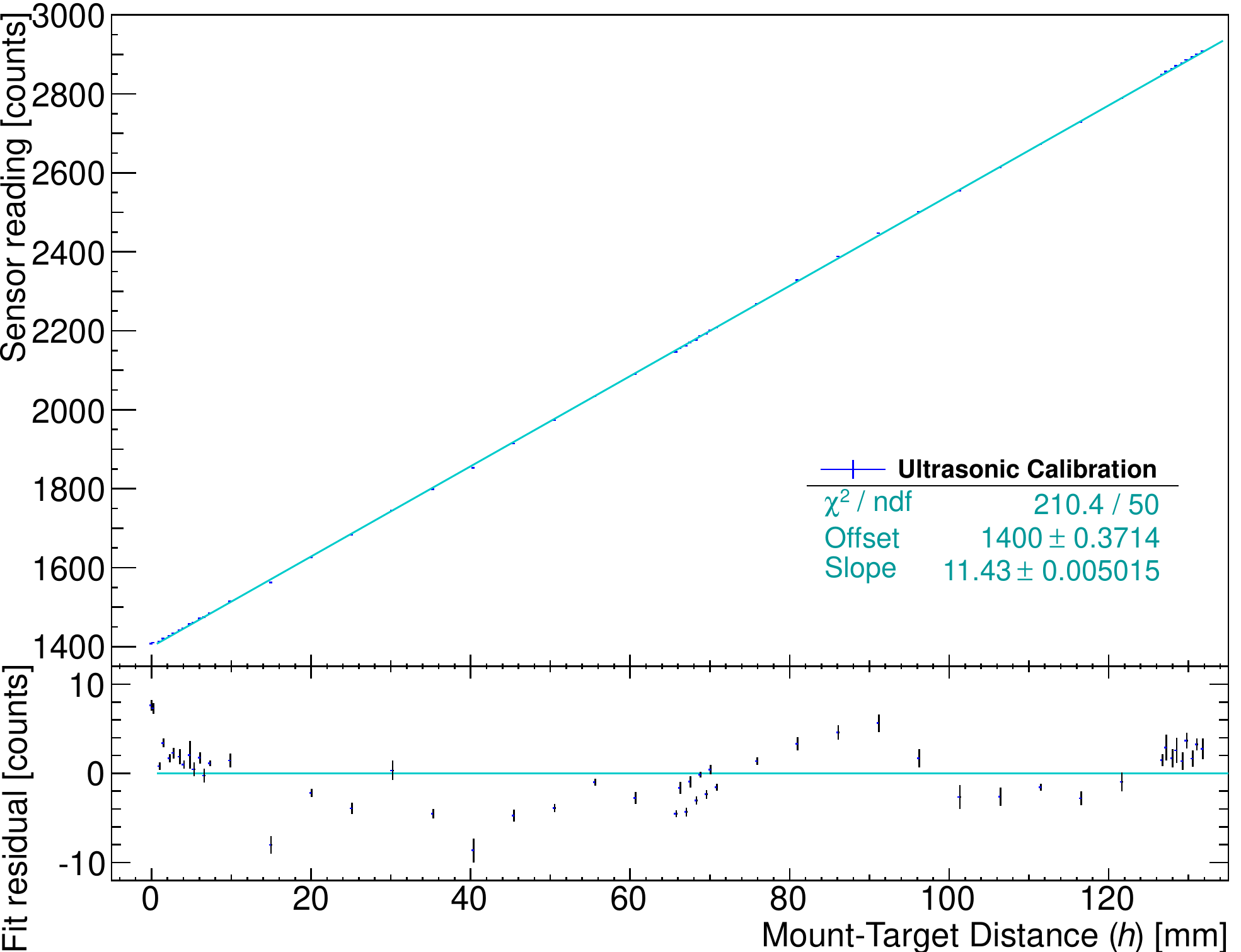}}\hfil%
\caption{The ultrasonic sensor calibration setup and one calibration result.}\label{fig:ultracalibration}
\end{figure}

The ultrasonic sensors were calibrated in a laboratory environment using the same apparatus diagrammed in Figure~\ref{fig:ultracalibrationstand}. Each sensor was set up as shown in Figure~\ref{fig:ultrasonic}. 
The ultrasonic pulse was directed out the bottom of the mount onto a target, a flat piece of acrylic. 
 The target was fixed on one jaw of an outside Vernier caliper, with the other jaw fixed on a laboratory optical bench. The upper jaw could be adjusted over a wide range, from touching the bottom of the sensor mount ($h=0$ in Figure~\ref{fig:ultracalibrationstand}) to about $400\unit{mm}$ ($15\unit{in}$) from the bottom of the mount. The target was positioned at several points within this range, and the caliper height recorded along with the sensor reading. Sample data from one ultrasonic sensor's calibration is shown in Figure~\ref{fig:ultracalibrationgraph}.

Because the distance from the face of the installed ultrasonic sensor to the bottom of the overflow tank is difficult to measure accurately with common measuring tools, the ultrasonic sensor is used to determine its own zero value. This method also avoids the problem of having to determine the offset between the true path length through the calibration apparatus of Figure~\ref{fig:ultracalibrationstand} and the caliper reading. With the sensor installed in its final position in the dry overflow tank, a baseline sensor reading was recorded. This reading defines the zero-fluid point for that tank. The calibration apparatus height range was chosen so that this point would always lie near the middle of the calibrated range. Thus, subsequent changes in the sensor reading caused by increasing fluid level can be accurately converted to an increase in fluid height with the calibration data.

The ultrasonic sensor measurements are especially sensitive to the speed of sound in the medium carrying the sound pulses. The calibration described here was performed in ordinary laboratory air, while the operating environment for the sensors is dry nitrogen gas. The difference between the speed of sound in nitrogen and in air is large enough that a correction was necessary. During sensor calibration, we recorded the temperature, pressure, and relative humidity of the laboratory air, then calculated the speed of sound using the expression provided by Cramer \cite{cramer1992}, with updated absolute humidity formulas from \cite{picard2008}. In operation, the speed of sound is calculated using the calibration equation of Span et al.\ for dry nitrogen \cite[\textsection7]{span2000}, incorporating the frequency correction of \cite[\textsection{4.1.2}]{span2000}. The cover nitrogen gas exchange is slow enough that the temperature measured by the liquid temperature sensors of section~\ref{sec:OFtempsensors} is approximately equal to the cover gas temperature. A nominal supply gas pressure of $21\unit{psia}$ is used for calculations. The variation of the speed of sound in nitrogen with pressure is small\footnote{In dry nitrogen at $22\degC$, the speed of sound changes by less than $200\unit{ppm}$ over the pressure range from $14$ to $21\unit{psia}$, a much larger variation than we expect in the detector cover gas supply pressure.
} and can be neglected. Overall, the correction due to the varying speed of sound is about a $3\unit{mm}$ change in the measured zero point of each overflow tank, or about a $1\unit{mm}$ change in the measured operating liquid level.

\begin{figure}[t]\centering
\includegraphics[width=75mm]{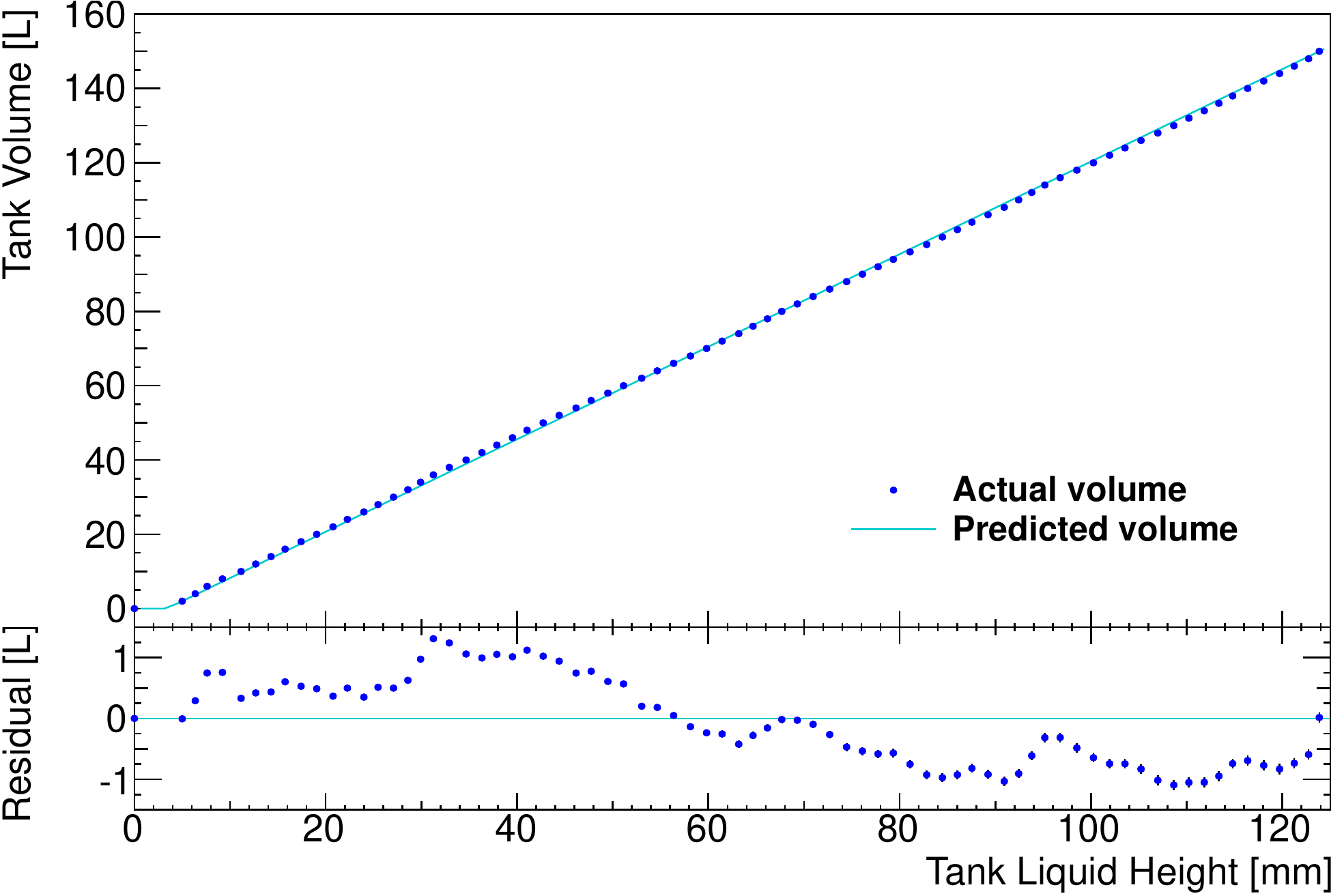}
\caption{Results of the overflow tank geometry and sensor performance cross-check. The largest observed deviation from the design geometry is $1.5\unit{L}$, which provides the ultimate maximum error bound on the overflow tank mass determination.}\label{fig:OFtankPSLgraph}
\end{figure}

As the ultrasonic sensors serve as the primary fluid level reference, an extra cross-check was performed. One overflow tank was set up in a laboratory environment and filled with high-purity water in $2000\unit{mL}$ increments, taking ultrasonic sensor readings at each step. (The capacitive sensors could not be cross-checked during this test, as they respond differently to water than to their operating liquids.) This test provided a full-system cross-check on both the response of the ultrasonic sensors and the geometry measurements of the overflow tanks. Data from the test is shown in Figure~\ref{fig:OFtankPSLgraph}. The maximum observed deviation from the predicted liquid level at a specified fluid volume was $1.5\unit{mm}$, which we take as the maximum height uncertainty of the liquid level monitoring system.

\subsection{Capacitive liquid level sensors}\label{sec:capcalibration}

\begin{figure}[t]\hfil
\subfloat[A schematic of the PTFE capacitive liquid level sensor calibration stand. The immersion depth $D$ of the sensor is calculated in terms of the spacer height $A$, stand height $B$, and sensor length $C$ as $D=C-(A+B)$.]{\label{fig:capcalibrationstand}\includegraphics[width=60mm,height=60mm]{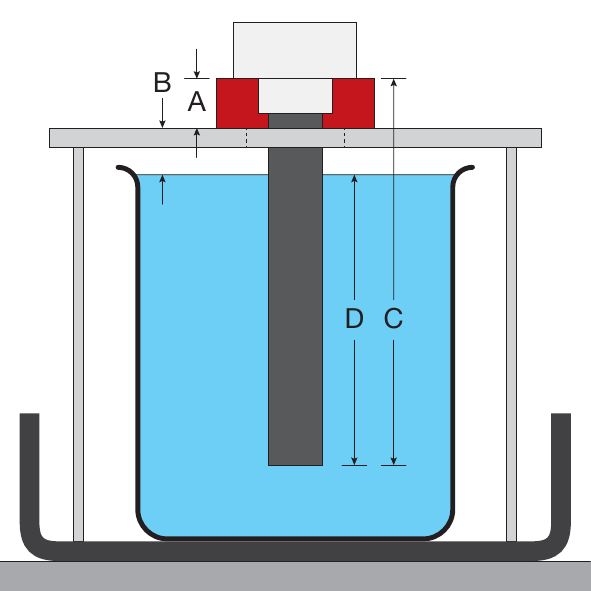}}\hfil
\subfloat[The results of sensor calibration for PTFE capacitive liquid level sensor 15, showing data from, fits to, and residuals for each run separately. For the combined fit, the two wet calibrations were averaged, excluding the dry calibration.]{\label{fig:capcalibrationgraph}\includegraphics[height=60mm]{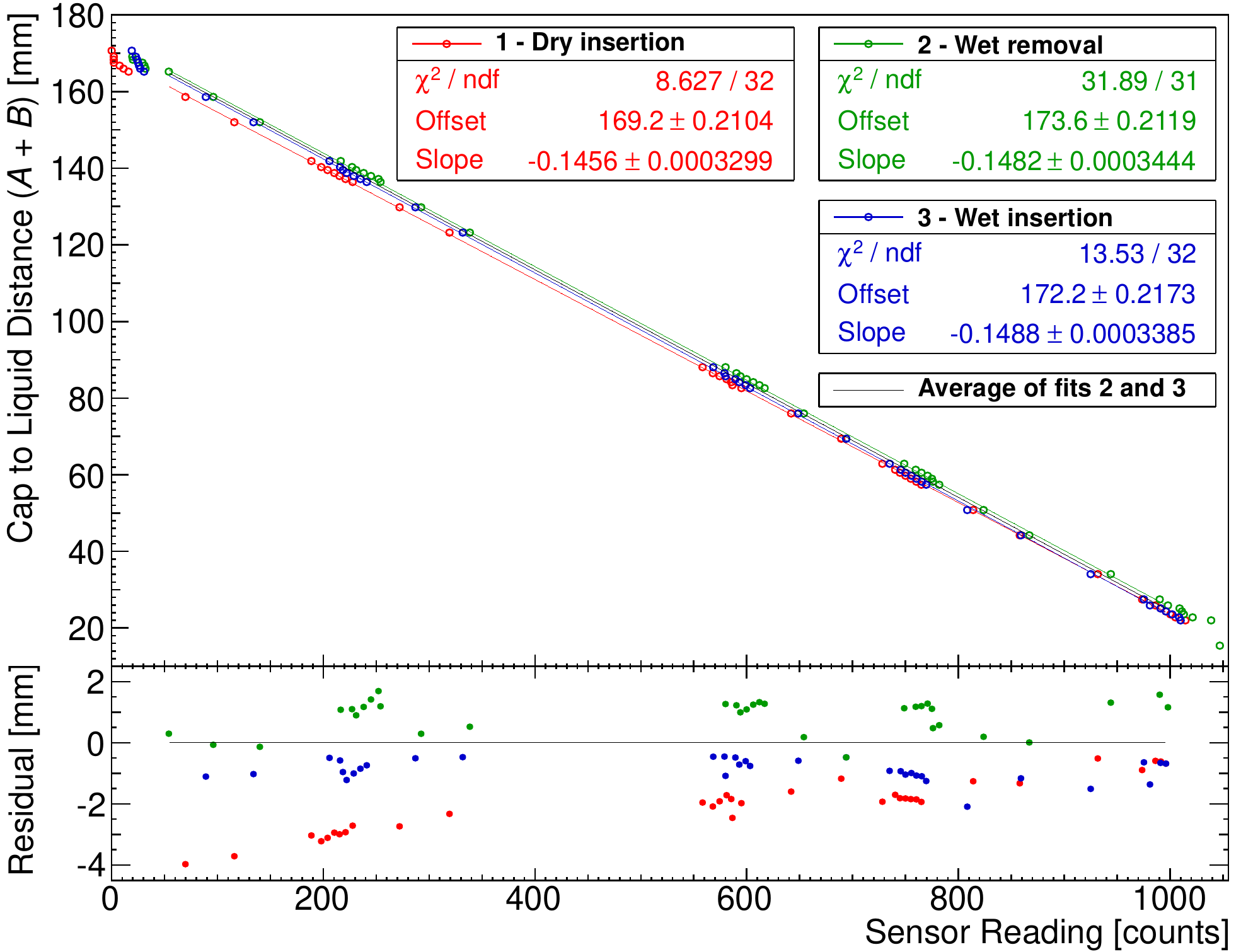}}\hfil%
\caption{The capacitive liquid level sensor calibration stand and one calibration result.}\label{fig:capcalibration}
\end{figure}

The GdLS and LS volume capacitive liquid level sensors were calibrated using the setup shown in Figure~\ref{fig:capcalibrationstand}. Each sensor was immersed in a beaker of linear alkyl benzene (LAB) in a spill tray. The beaker was kept continually full to the brim with LAB, topped off as necessary; liquid spilled out of the beaker at the same height, so topping off the LAB until the point where liquid spilled out ensured a constant liquid level in the beaker (i.e., dimension $B$ in Figure~\ref{fig:capcalibrationstand} remained fixed). The sensor height was adjusted by changing the height of spacers, shown in Figure~\ref{fig:capcalibrationstand} in red, with thickness given by dimension $A$. Spacer height ranged from a minimum of zero (no spacers, sensor bottom flush with support beam) to approximately $160\unit{mm}$, slightly greater than the total length of the sensor barrel, leaving the tip of the sensor suspended over the beaker. The spacers used were thin U-shaped aluminum shims for fine adjustments and aluminum tubes for larger adjustments. 

Each sensor was calibrated in three passes. Before the first pass, the sensor was cleaned in pure ethanol and left to dry; ethanol and LAB are miscible, so this rinse left the sensor in a clean state. The dry sensor was then lowered into the beaker of LAB, decreasing dimension $A$ in Figure~\ref{fig:capcalibrationstand} as the sensor descended. After reaching the bottom, the sensor was then lifted out during the second pass. The beaker was kept continually topped up during this pass to maintain a constant liquid level. The third pass repeated the first pass, but with the sensor now fully wetted with LAB. All three passes were recorded and analyzed to calibrate the sensor readout. A sample calibration result from one sensor is shown in Figure~\ref{fig:capcalibrationgraph}. The final calibration result used the average of the two wetted runs, excluding the dry calibration result. During operation we expect small level changes over long time scales, which corresponds more closely to the performance of an already immersed sensor.

The calibration of the stainless steel mineral oil capacitive level sensors followed a similar procedure. A much larger quantity of mineral oil was available, so all nine mineral oil sensors were calibrated together in one large tank of oil. The oil level in this tank was directly controlled, avoiding the use of spacers or any adjustment of the sensors. Only one run was performed, starting with a dry sensor and finishing with a fully immersed sensor. Multiple calibration runs were not performed as one run was sufficient to achieve our accuracy goals for mineral oil level monitoring.

Due to the tolerances of the sensor mounts, the capacitive level sensor calibration alone was not enough to establish the absolute liquid level in the central overflow volumes. After each overflow volume was filled with liquid to its operating level, the absolute reading from the ultrasonic sensor in that volume was used as a cross-reference to establish the overall offset of the capacitive sensor. While this approach is not as robust as an absolute determination of the liquid level using only the capacitive sensor, it still allows the capacitive sensor to serve as a secondary level monitor, checking any drifts of the ultrasonic sensors and providing redundancy in case of sensor failure.

Cross-referencing is not necessary for the mineral oil sensors, which have an adjustable mount. The installation procedure for the mineral oil sensors uses a model capacitive sensor that is $2.0 \unit{mm}$ longer than the real sensors. The model sensor is installed, the sensor mount is adjusted until the bottom of the model sensor just touches the bottom of the mineral oil overflow tank, then the model is replaced with the real sensor. This procedure establishes the calibrated sensor reading as the tank liquid level minus $2.0\unit{mm}$.

\subsection{Overflow temperature sensors}

\begin{figure}[tb]\centering
\includegraphics[height=50.mm]{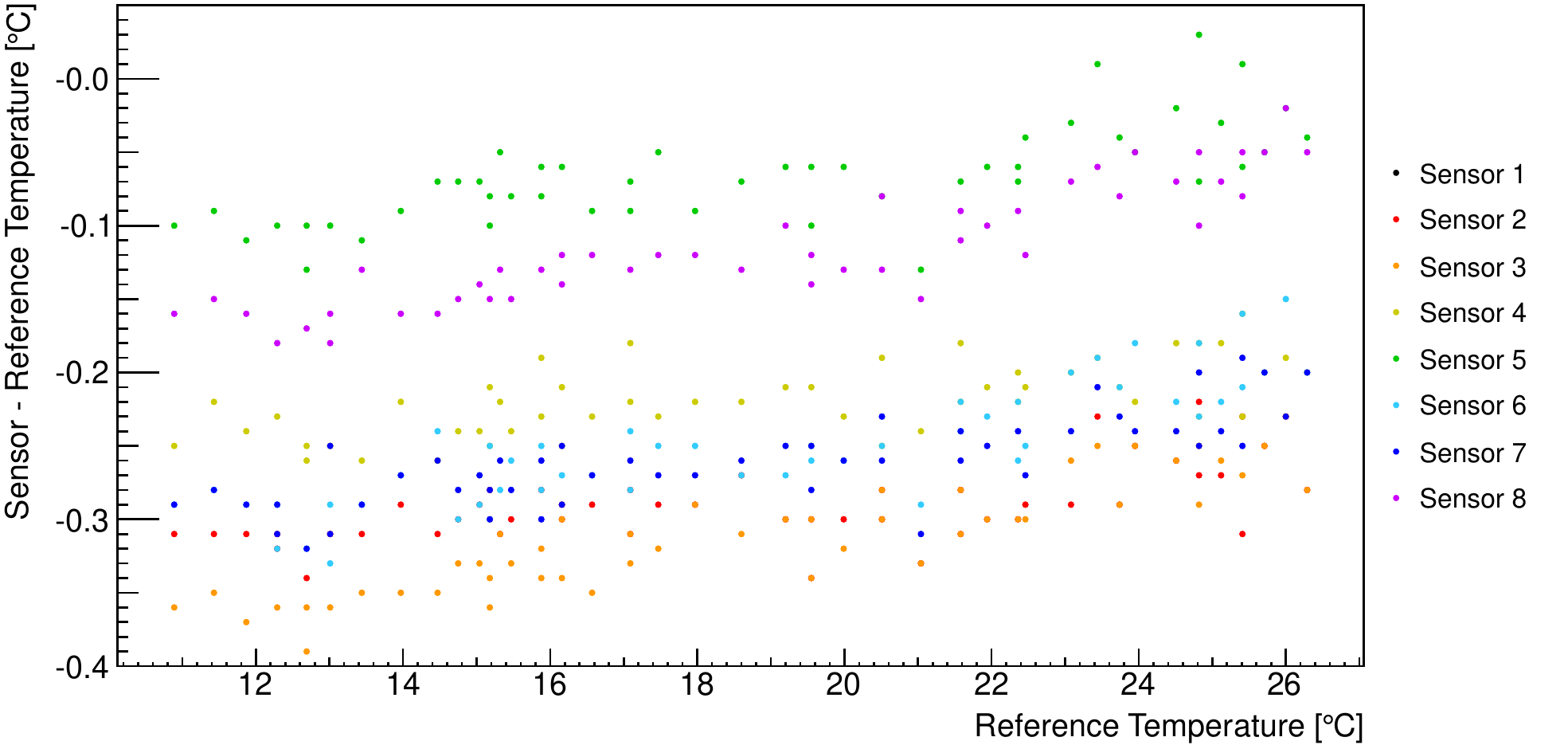}
\caption{Calibration data from the overflow temperature sensors installed in ADs 5--8, all of which were calibrated together. To better illustrate the sensor behavior, only the deviations from the reference reading are shown. The spread in uncalibrated readings of up to $0.4\degC$ is negligible after calibration.}\label{fig:tempcalibration}
\end{figure}

The overflow tank temperature sensors were calibrated in water baths. The Pt100 sensors were placed in a water bath along with a NIST-traceable reference thermometer\footnote{Omega Engineering model HH41 with ON-410-PP temperature probe, serial number 308697, system calibration ID 009968179.}. We performed a full-system calibration: each sensor was paired with its electrical readout module channel before calibration, and this pairing was maintained at sensor installation. Sensors were calibrated in batches. Groups of four or eight Pt100 sensors were placed in the water bath together, comprising complete sets of overflow temperature sensors for two or four detectors, respectively. Data was first taken at room temperature, then ice was added to the water bath, and several data points were taken as the system cooled down to $0\degC$. The water bath was frequently agitated to reduce thermal gradients in the cooling water. Calibration results from a batch of eight sensors calibrated together are shown in Figure~\ref{fig:tempcalibration}. All sensors exhibited good linearity within and beyond the detector's operational temperature range, meeting the requirements of section~\ref{sec:requirements}.

\begin{figure}[tb]\hfil%
\subfloat[An inclinometer mounted on the central overflow tank.]{\label{fig:inclinometerphoto}\includegraphics[clip=true,trim=9in 6in 0 6in, width=74mm]{figures/inclinometer_mounted.jpg}}\hfill%
\subfloat[The calibration data from inclinometer 16, showing the measurement of the sensor's response and zero offset.]{\label{fig:tiltcalibration}\includegraphics[width=74mm]{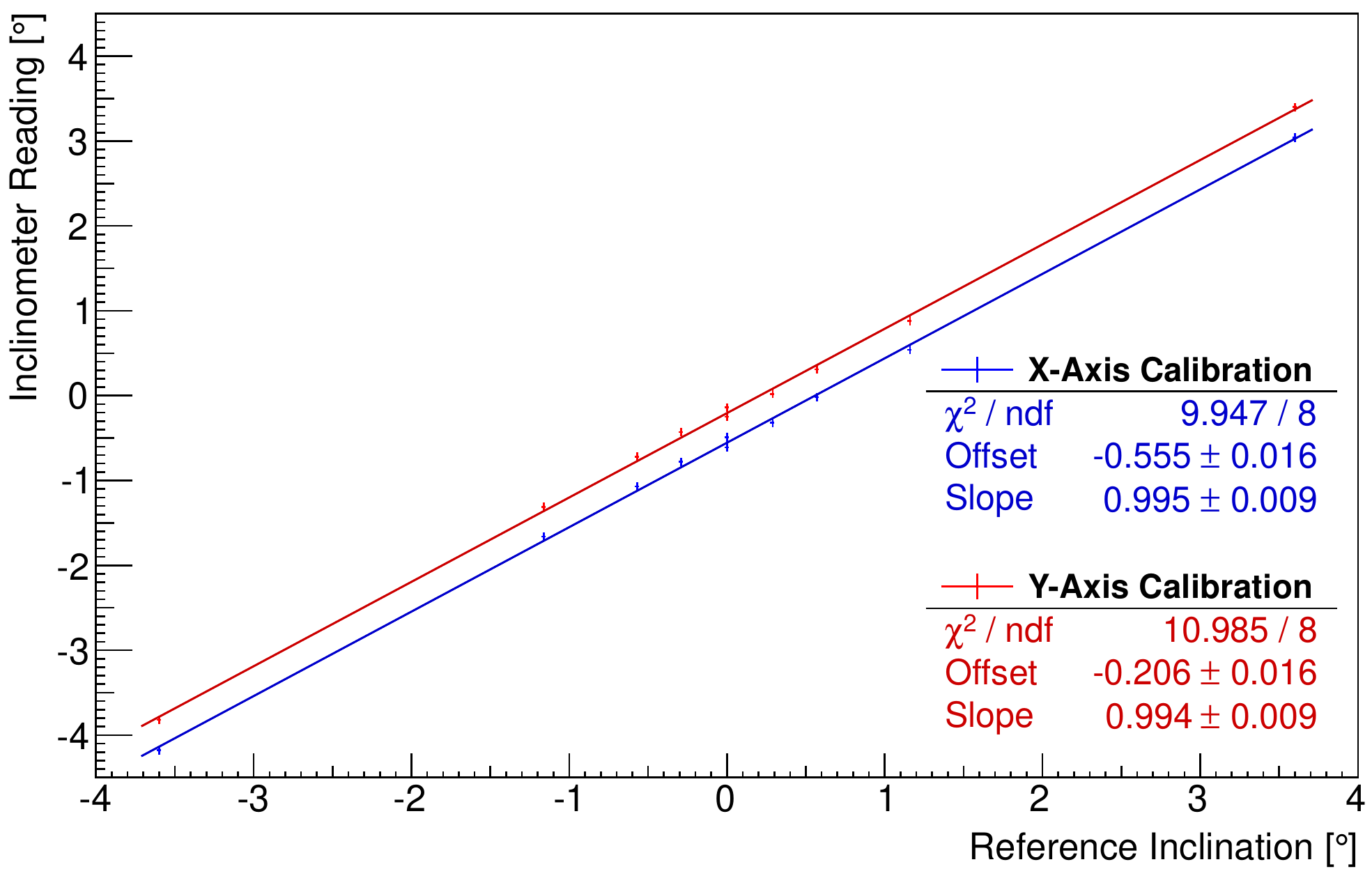}}\hfil%
\caption{Images of the inclinometers and a sample of the inclinometer calibration data.}
\end{figure}

\subsection{Inclinometers}

The inclinometers also were calibrated in a laboratory setting before installation. The inclinometers are digital sensors and directly report the measured deviation of their $x$ and $y$ axes from the plane perpendicular to local gravity. We determined the zero offset of each inclinometer by taking a reading on a known-level surface, as checked to within $0.05\deg$ using a SPI-Tronic Pro 3600 digital level. We then used a custom-fabricated 150-{mm} sine bar to tilt each axis of each inclinometer by a known amount and tabulated the readings. Sample results from this calibration are shown in Figure~\ref{fig:tiltcalibration}, demonstrating that all sensors tested showed satisfactory zero readings and inclination response.

\section{Instrumentation readout}
\subsection{Hardware design}

\begin{figure}[t]\centering
\includegraphics[width=150mm, height=75mm]{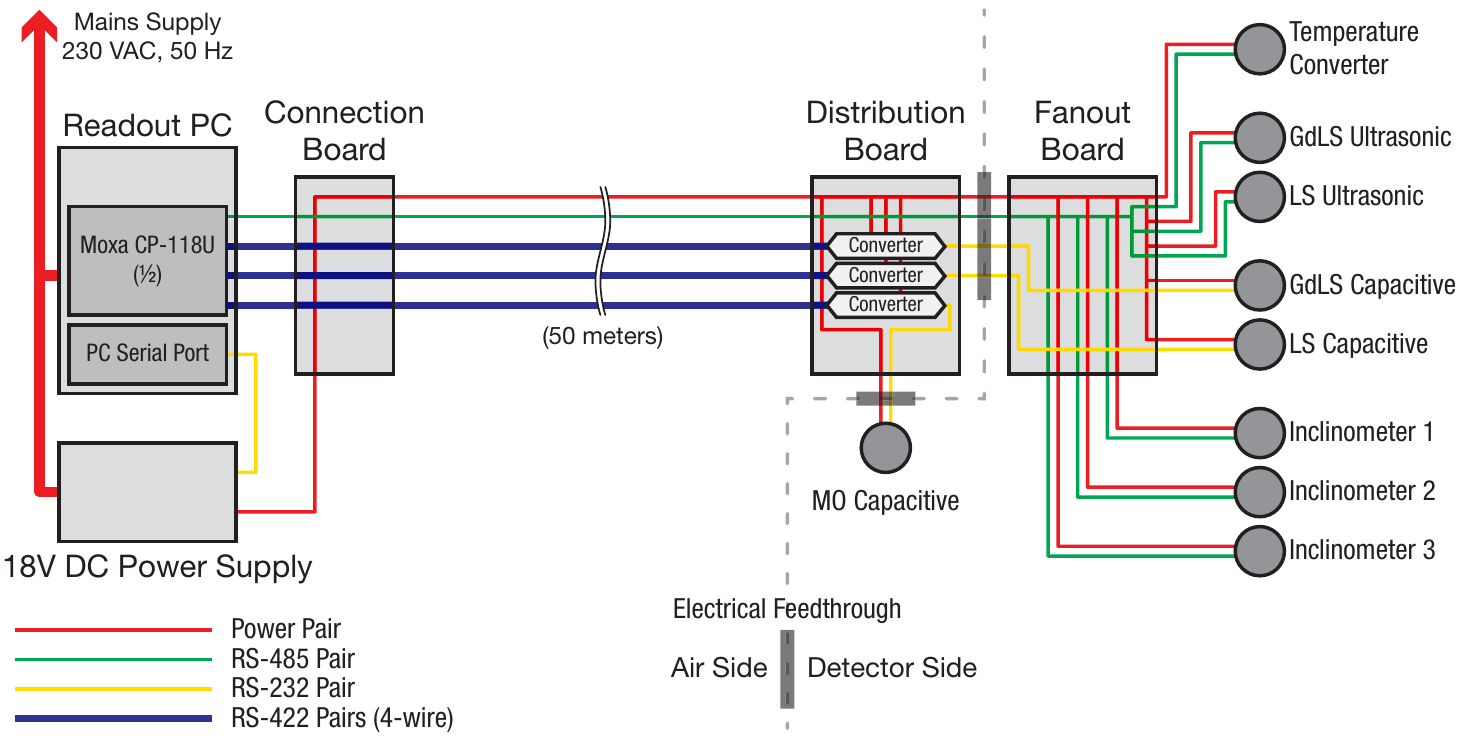}
\caption{Electrical block diagram of the lid sensor system. Line color indicates  signal type; the boundary between the air side of the system and the detector side is shown by a dashed line, with electrical feedthroughs in grey. Some power lines shown as single pairs are multiple parallel conductors for improved redundancy and increased capacity. All cabling is shielded.}\label{fig:lidsensorblockdiagram}
\end{figure}

The electrical layout of the main lid sensor system is shown in Figure~\ref{fig:lidsensorblockdiagram}. The sensors are all physically located within the detector volume, which is sealed and leak-tight against the water pool surrounding the detector. All sensors are connected through serial interfaces to a readout computer, which has a PCI multi-port serial card installed (a Moxa CP-118U, with eight ports each independently configurable for RS-232, RS-422, or RS-485 communication). Each detector requires four of the eight ports; one readout computer can read out two detectors with one serial card, or four detectors using two cards. The readout software, written in National Instruments LabVIEW 8.6.1f1 for compatibility with other detector readout and control systems, polls each sensor 
every two seconds, redundantly recording the data into both a flat log file and the experiment's SQL database. The flat log file is easy to use for manual checks or system debugging, while the database provides the measured data directly to the experiment's analysis pipeline. A single DC power supply with a serial port connection provides a stable, continuously-monitored, remote-controllable power source for all sensor systems connected to a readout computer.

The 50-meter distance between the detector-side interface and the readout computer means that conventional serial interfaces, such as USB or RS-232, cannot be used. Instead, differential electrical interfaces are used on the main connection between the detector and the readout computer for all sensors. Most sensors were available in RS-485 variants. The 2-wire, half-duplex operation of RS-485 is well suited to low-speed, low-volume data readouts like our sensors, and is capable of line lengths of much longer than $50\unit{meters}$ even in a noisy electrical environment. One common RS-485 line services all RS-485 sensors. The line is run directly from the readout computer to the fanout board on the detector lid, where it switches to a star topology to connect to each sensor. The use of a star topology rather than the preferred daisy chain was necessitated by the practical concerns of cable routing. The short lengths of the star branches, one to two meters each, do not cause interference issues, as the signal travel time is much smaller than the typical symbol length at our data rate of 9600 baud.

The capacitive liquid level sensors were not available with an RS-485 interface, only the more common RS-232. To carry these signals to the readout computer, they are first converted to RS-422, a four-wire, full-duplex, differentially-signaled interface, logically compatible with RS-232. The communication protocol used by the capacitive sensors requires a full-duplex link, necessitating the use of RS-422 and a dedicated line for each sensor. The signal converters are situated on the electrical distribution board, mounted on the detector lid just outside the sealed central overflow volume, and powered by the common power bus. This arrangement limits the length of the less-reliable RS-232 link to only a few meters, using RS-422 for the majority of the line length.

With the exception of the temperature readout, all sensors on the common RS-485 line use checksummed communications protocols, such as Modbus/RTU, so each sensor ignores commands not directly addressed to it with matching checksums. The HWg-Poseidon protocol used by the temperature sensor does not have a checksum \cite{HWgPt100}. We have seen infrequent readout glitches caused by the temperature sensor misinterpreting commands to, or responses from, other sensors as a request to read out, causing a transient bus conflict that lasts until the temperature sensor stops transmitting. Conflicts are rare in practice, but force our readout system and data analysis chain to be able to handle a failed reading without complications. The use of a checksummed protocol for all sensors, or the use of dedicated per-sensor lines as  with the capacitive sensors, would have eliminated this issue.

\begin{figure}[tb]\centering\includegraphics[width=150mm]{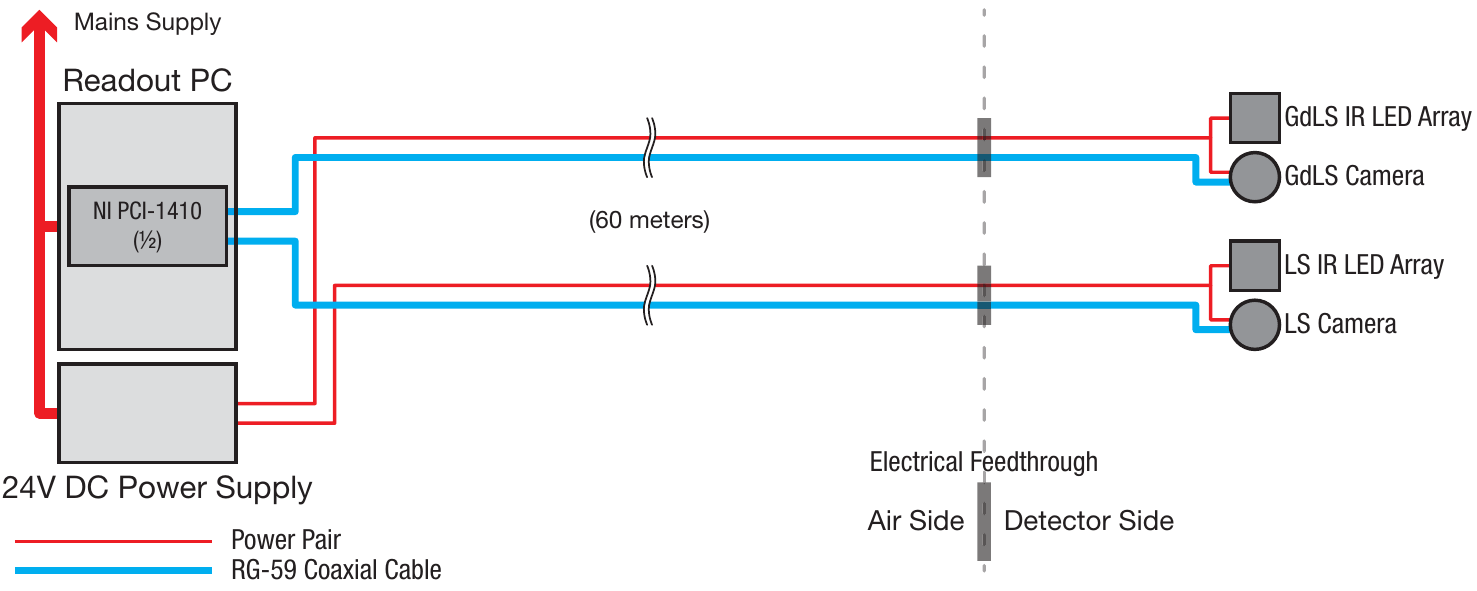}\caption{Electrical block diagram of the off-center camera system. Line color indicates  signal type; the boundary between the air side of the system and the detector side is shown by a dashed line, with electrical feedthroughs in grey.}\label{fig:DGUTcamerablockdiagram}\end{figure}

The electrical layout of the off-center camera system is shown in Figure~\ref{fig:DGUTcamerablockdiagram}. As with the lid sensors, each readout system is able to service one pair of detectors, with two cameras per detector. The cameras transmit analog PAL video signals back to the readout computer over RG59 coaxial cables. A National Instruments PCI-1410 PCI card in the readout computer receives and digitizes the video signals. Each PCI-1410 has four analog video inputs. One twisted pair carries $24\unit{V}$ DC power to each camera system and the infrared LED array. The power supply is shared between all cameras connected to the same readout PC.

\subsection{Long-term detector monitoring}

\begin{figure}[t]
\centering
\includegraphics[width=\textwidth]{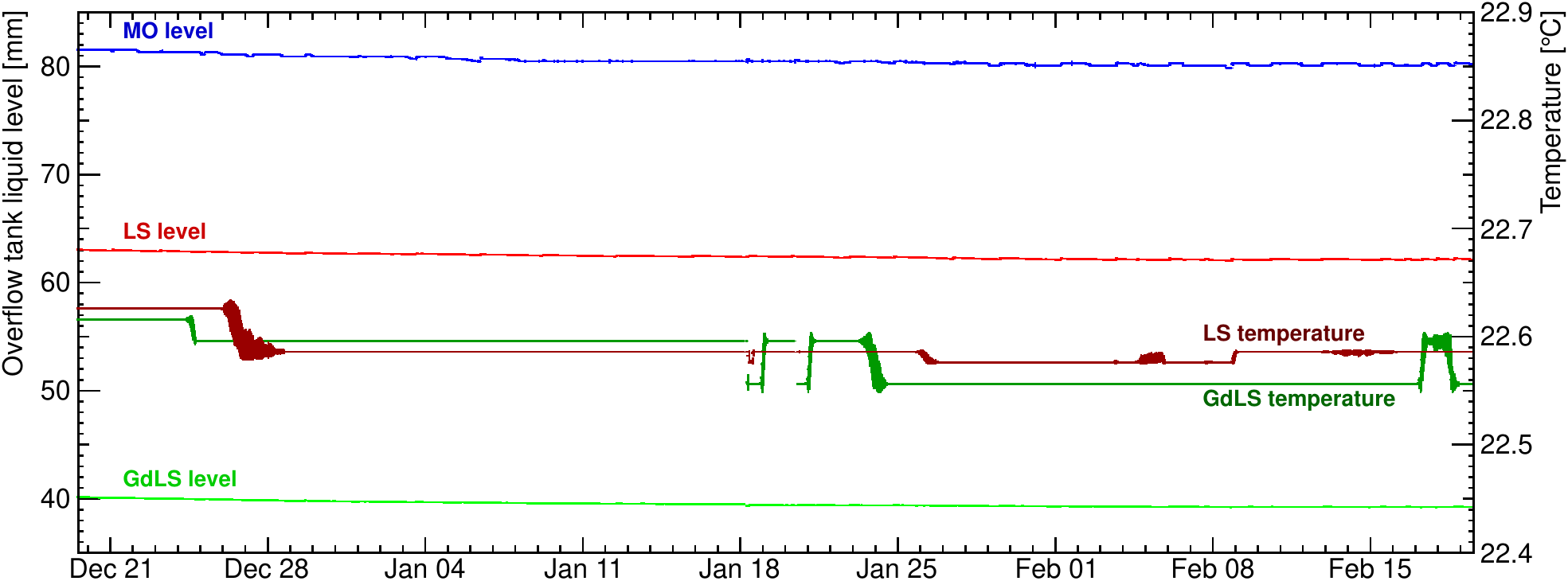}\caption{Lid sensor monitoring data from AD1 during the first physics run, using data from 24 December 2011 to 17 February 2012. Overflow tank GdLS, LS, and MO levels are shown on the left scale, with GdLS and LS temperatures on the right scale. Data points are hourly averages.}
\label{fig:monitoringdata}
\end{figure}

\begin{figure}[t]\hfil
\subfloat[The GdLS liquid level and temperature in AD3. Arrows indicate the times the pictures at right were taken.]{\includegraphics[width=75mm]{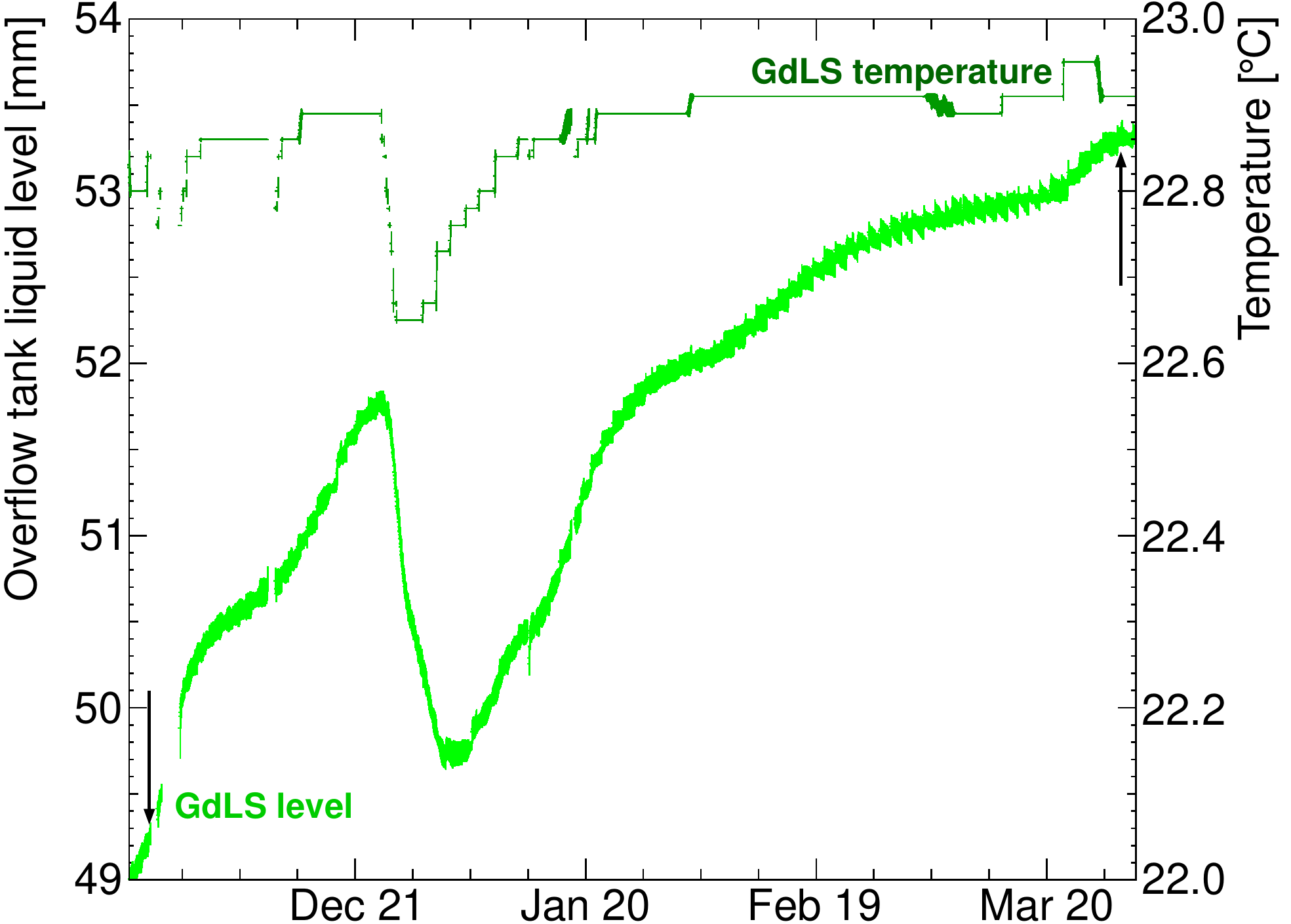}}\hfill
\subfloat[A side-by-side image showing GdLS level changes in AD3, from 24 Nov 2011 (left) to 30 Mar 2012 (right).]{\includegraphics[width=73mm]{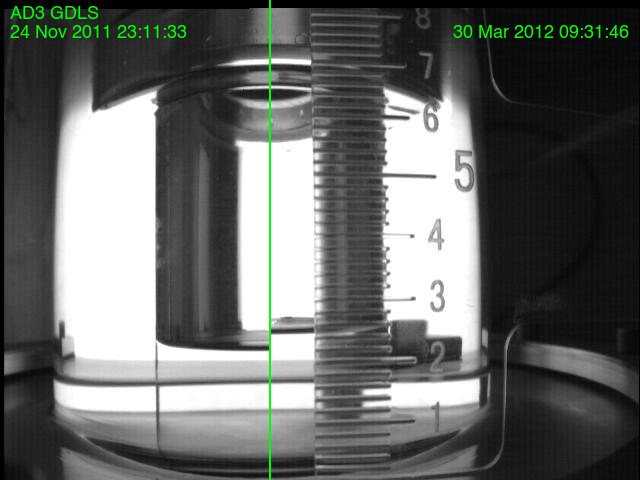}}\hfil%
\caption{GdLS levels in AD3 before, during, and after the first physics run. The camera and the level sensor both show the same change of approximately $4\unit{mm}$.}\label{fig:cameracomparisonshot}
\end{figure}

The overflow monitoring sensors have been operated continuously since the completion of detector installation, recording sensor readings once every two seconds.  A sample of the data from AD1 installed at the Daya Bay near site is shown in Figure~\ref{fig:monitoringdata}. A split camera image from AD3 is shown in Figure~\ref{fig:cameracomparisonshot} together with sensor data. In all detectors, the liquid level sensors show the same behavior as the camera images. Camera images are taken at least once per month during scheduled detector calibration periods when physics data is not being collected.


A long-term laboratory stability test is underway with a full set of sensors, production readout system, and environmental monitor installed on a test bench at the University of Wisconsin--Madison. This setup allows us to monitor the sensors for any long-term drift unrelated to the behavior of the fluids. The sensors are readily accessible, so if any future drifts in sensor readings are observed, both the sensors and the test environment can be investigated to determine the cause. We expect to operate this long-term test stand for at least the duration of the first long-term physics run at Daya Bay. 

\subsection{Monitoring during detector transport}

\begin{figure}[bt]
\centering
\includegraphics[width=\textwidth]{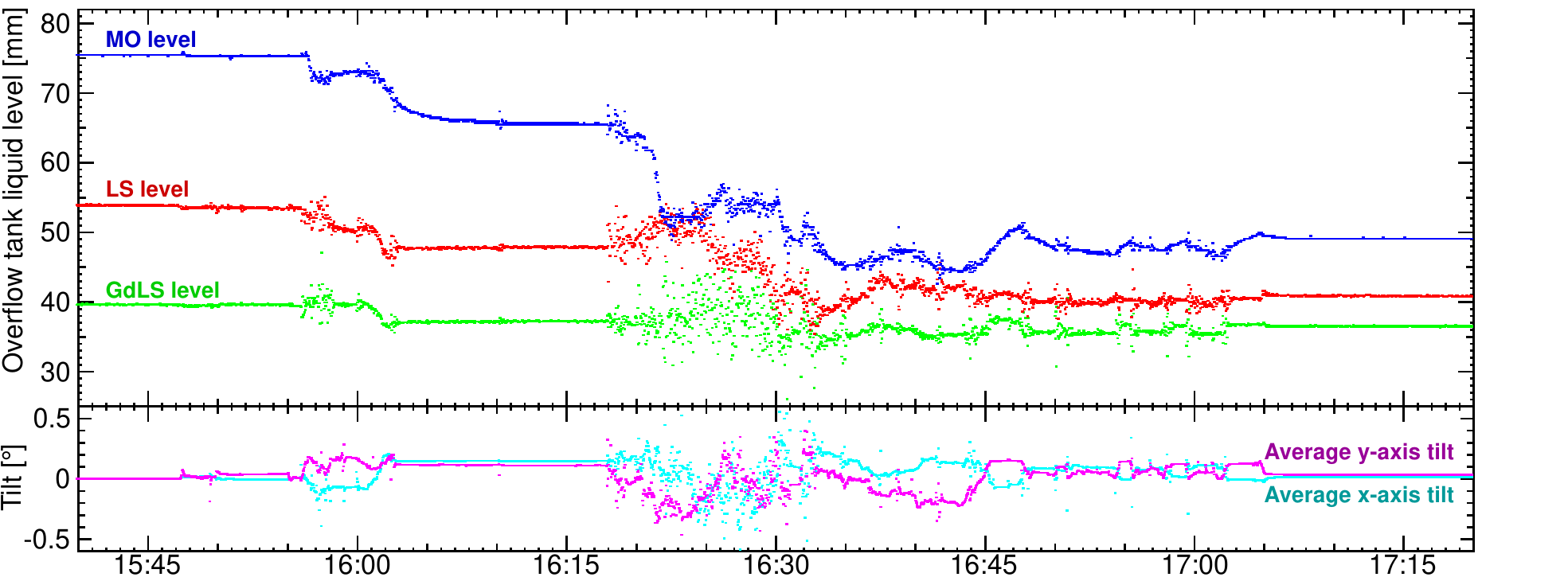}
\caption{Overflow sensor data collected during transportation of AD1 from the detector filling hall to Experimental Hall~1 on 11 May 2011. Times are Daya Bay local time, UTC~+0800. For clarity, the raw level sensor readings are displayed without error bars. Inclinometer values shown are the average of all three inclinometer readings, zero-corrected to show change from initial state.}
\label{fig:AD1transportdata}
\end{figure}

After assembly and filling, the Daya Bay ADs were transported to their final locations in the experimental halls. When full, an AD weighs about $110\unit{tons}$ and can only be transported underground using an automatic guided vehicle (AGV), a low-profile heavy-lift transport capable of bearing and moving the entire load~\cite{agv}. As one check to verify that the ADs were not damaged during transport, we ran the liquid level monitoring instrumentation and monitoring cameras during the first transport of AD1 to its installed location. A readout PC and supporting instrumentation were mounted in a mobile 19-inch rack with an uninterruptible power supply battery system to provide power; this rack trailed behind the AGV as it transported AD1 and allowed us to monitor AD1 liquid levels in real time during transport.

The data collected during the transport of AD1 is shown in Figure~\ref{fig:AD1transportdata}. Several notable features can be seen in this data. The measured liquid levels in each volume decreased, which we attribute to the agitation of the detector dislodging trapped bubbles of gas. The inclinometers show that the detector remained level within the expected range during the transport. Based on the stability of AD1 during transport, we decided not to monitor future ADs continuously during transportation, instead checking their liquid levels before and after transportation. The results have been consistent with AD1: we see no overall change except for a decrease in liquid levels associated with the release of trapped gas bubbles. We also performed a final top-off of all detector liquids after detector transportation\citefillingpaper, to replace the volume lost when the gas  bubbles were released.

\section{Summary}
We present a comprehensive and redundant system of sensors for monitoring the volume, mass, and temperatures of the mineral oil (MO), liquid scintillator (LS), and gadolinium-doped liquid scintillator (GdLS) inside the Daya Bay antineutrino detectors (ADs).
The GdLS and LS fluid levels are monitored continuously by ultrasonic and carbon-impregnated PTFE capacitive sensors in the liquid overflow tanks, and are cross-checked by off-center cameras next to the detector calibration tubes.  The MO fluid level is monitored by a stainless steel capacitive sensor.  Two temperature probes monitor the GdLS and LS fluids, respectively, and four temperature probes evenly-spaced vertically along a photomultiplier tube support ladder read the MO temperature and determine the vertical temperature profile of the detectors.  In addition to the fluid monitoring sensors, three inclinometers monitor the $x$- and $y$-axis tilts of the AD, and two internal cameras allow visual inspection of the inside of the AD.


Uncertainty of the target mass inside the detector is potentially one of the largest systematic uncertainties on a measurement of $\sin^2 2\theta_{13}$.  
This system of sensors monitors the height of the target liquid in the overflow tanks to a precision of $1\unit{mm}$. When combined with the overflow tank geometry information, GdLS density measurements, and tilt sensor data, the set of sensors described here is able to track detector target mass changes with a maximum uncertainty of $2.2\unit{kg}$ (0.011\%), significantly less than the baseline design goal of $4\unit{kg}$ (0.02\%). Minimizing this uncertainty has played a key role in reducing the systematic uncertainty of Daya Bay's measurement of $\sin^2 2\theta_{13}$~\cite{DYBcpc}.

\section*{Acknowledgements}
This work was performed with support from the DOE Office of Science, High Energy Physics, under contract DE-FG02-95ER40896 and with support from the University of Wisconsin.

The authors would like to thank 
B. Broerman and E. Draeger for assistance with the overflow tank assembly process;
A. Green, P. Mende, J. Nims, D. Passmore, and R. Zhao for assistance with the overflow sensor characterization and calibration;
Z.\,M. Wang for assistance with the off-center camera system surveys;
X.\,N. Li for invaluable assistance on-site;
and
C. Zhang for integrating our systems into the online data monitor.
A. Pagac and B. Broerman helped prepare many of the figures in this paper.
Many fruitful discussions with the rest of the Daya Bay Collaboration have helped to refine the systems discussed here and contributed greatly to their excellent performance.

\bibliographystyle{JHEP}
\bibliography{ref}

\end{document}